\documentclass[12pt]{article}
\usepackage{graphicx}
\usepackage{geometry}
\usepackage{amssymb}
\usepackage{setspace}
\usepackage[dvipsnames]{pstricks}
\usepackage{pst-node}
\usepackage{pst-slpe}
\usepackage{pst-tree}
\usepackage{bbm}
\usepackage{amsmath, amsthm, amssymb}
\usepackage{pgfplots}
    \pgfplotsset{width=8cm,compat=1.5.1}
\usetikzlibrary{arrows.meta}
    \usetikzlibrary{shapes,snakes}
\usepackage{amsmath,amssymb,amsthm}
\usepackage{tikz}
\usepackage{caption}
\usepackage{subcaption}
\usepackage{booktabs}
\usepackage{empheq}
\usepackage[most]{tcolorbox}
\usepackage{diagbox}
\usepackage{blkarray}

\addtocontents{toc}{\protect\setcounter{tocdepth}{1}}
\appto\appendix{\addtocontents{toc}{\protect\setcounter{tocdepth}{1}}}

\newtcbox{\mymath}[1][]{%
    nobeforeafter, math upper, tcbox raise base,
    enhanced, colframe=blue,
    colback=yellow!30, boxrule=1pt,
    #1}
    
\usepackage{natbib}
\usepackage{draftwatermark}
\SetWatermarkText{}
\SetWatermarkScale{4}

\newtheorem{theorem}{Theorem}

\newtheorem{axiom}{Axiom}

\newtheorem{example}{Example}

\newtheorem{lemma}{Lemma}

\newtheorem{observation}{Observation}

\newtheorem{proposition}{Proposition}

\newtheorem{definition}{Definition}
\usepackage{tikz}
\usepackage{dirtytalk}

\title{Intertemporal Aggregation of Choice Data with Consumption Dependent Preferences
\footnote{I am grateful to Victor Aguiar, Roy Allen, Miguel Ballester, Laurent Bouton, Simone Cerreia-Vioglio, Edward Honda, Sean Horan, Roger Lagunoff, Jay Lu, Andrew Mackenzie, Kyle Monk, Tianshi Mu, Collin B. Raymond, John Rehbeck, Marciano Siniscalchi, and Joshua Teitelbaum  as well as seminar participants at Bocconi, BRIC 2023, Bristol, Georgetown, NASMES 2023, Queensland, RUD 2024, and SAET 2023 for their helpful comments during the course of this project. I am especially grateful to Peter Caradonna, Christopher Chambers, and Yusufcan Masatlioglu for their continued support and insightful conversations throughout the course of this project. Much of the content of this paper was circulated previously in my job market paper ``Random Utility, Repeated Choice, and Consumption Dependence". \\
Turansick: Department of Decision Sciences and IGIER, Universit\'{a} Bocconi.  E-mail:  \texttt{christopher.turansick@unibocconi.it}}}
\author{Christopher Turansick}
\date{\today}

\begin{document}

\onehalfspacing

\maketitle

\begin{abstract}
    We study consumption dependence in the context of random utility and repeated choice. We show that, in the presence of consumption dependence, the random utility model is a misspecified model of repeated rational choice. This misspecification leads to biased estimators and failures of standard random utility axioms. We characterize exactly when and by how much the random utility model is misspecified when utilities are consumption dependent.
\end{abstract}

\section{Introduction}\label{Intro}

Random utility is a standard model of discrete choice. It is typically used to model either a population of heterogeneous but (stochastically) rational agents or to model the repeated choices of a single agent whose preference varies over time. Consumption dependence is the idea that an agent's history of choices impacts their utility today. Until now, it was not well understood how the presence of consumption dependence interacts with the repeated choice interpretation of the random utility model. We study exactly this and show that, in the presence of consumption dependence, the random utility model is a misspecified model of repeated rational choice.

One of the original goals of the random utility model was to explain the observation that agents vary their choice when repeatedly faced with the same environment. Random utility allows agents to be classically rational subject to a state or characteristics unobserved by the analyst. A key assumption of the random utility model is that the distribution over unobservables, and thus the distribution over preferences, is independent of the menu faced by the agent. Now suppose that an agent's history of choice impacts their preference today. It is certainly the case that the set of goods available to the agent, and by extension the set of goods potentially in their history of choice, can impact the realization of their preference today. We are not the first to make this observation, but we use the observation of \citet{machina1985stochastic} and others to motivate our work.
\begin{quote}
    ``While the random preferences approach seems a very natural explanation of individual variability, it nevertheless still possesses several troubling aspects. Does each choice situation induce a new realisation of the random preference ranking or do such realisations occur independently of the frequency of choice situations? Are successive realisations of the preference ranking independent of past realisations? of past choices? Is the realisation of the preference ranking at the time of a given choice situation stochastically independent of the particular set of alternatives available at the time, or if not, what is the nature of the dependence?" \citep{machina1985stochastic}
\end{quote}

In this paper, we tackle many of the questions posed in this quote by explicitly modeling a dynamic preference realization process where today's realization can depend on past preferences and choices. While our analysis is focused on the repeated choices of a single agent, the implications of our results have a wider reach. In discrete choice settings, within agent variation and between agent variation are both often modeled through random utility. As such, the dynamic implications of repeated choice are often ignored when aggregating across a population. Our motivating example is the market level analysis frequently done in empirical industrial organization. Consider \citet{nevo2001measuring} which studies cereal choice using market level data over four years and \citet{miller2017understanding} which studies beer choice using market level data over ten years. A typical consumer will face the cereal and beer consumption decision repeatedly and frequently. Despite this, each of these influential papers use a static specification of (random) utility and aggregate their data to the quarterly level.\footnote{Some specifications in these papers also consider aggregation of their data to the monthly level. Both papers use time period fixed effects in an attempt to capture dynamic heterogeneity of their static random utility model. We will discuss later how allowing for time period fixed effects may still allow for problems.} We later show that ignoring consumption dependence in such situations can lead to parameter estimates which are biased in their cardinal and ordinal predictions.

The misspecification and bias caused by consumption dependence is only as important as the prevalence of consumption dependence. Habit formation is a type of consumption dependence that supposes an agent is more likely to consume the good they just consumed. There is a long literature in macroeconomics which studies the prevalence of habit formation \citep{carrasco2005consumption} and its role in growth \citep{carroll2000saving} and monetary policy \citep{fuhrer2000habit}. As another example, reference dependence with status quo bias is a type of consumption dependence which asks that an agent's most recent choice acts as a reference point for their current choice. It has been well documented that agents are subject to status quo bias \citep{samuelson1988status,hardie1993modeling} and a long theoretical literature has sought to understand the empirical content of status quo bias \citep{masatlioglu2005rational,li2023random,kibrisrandom} and, more generally, reference dependence \citep{tversky1991loss,koszegi2006model,kovach2021reference}. As one last example, cognitive dissonance is a form of consumption dependence which says that an agent will change their preference to rationalize their past choices. There is a long literature in psychology showing that agents behave in a way consistent with cognitive dissonance \citep{harmon1999cognitive,chen2010choice}. All of this is to say that consumption dependence of various forms has been studied in various fields and has been shown to be prevalent in the decision making process.

\begin{figure}
    \centering
    \tikzset{every loop/.style={min distance=10mm,in=120,out=60,looseness=10}}
\begin{tikzpicture}[->,line width=1.25pt,pref/.style={ellipse, draw=blue!60, fill=blue!5, very thick, minimum size=7mm}]

        \node[pref,align=center] (a) at (0,3) {$a \succ c \succ b$};
        \node[pref,align=center] (b) at (0,0) {$b \succ c \succ a$};
        \node[pref,align=center] (c) at (6,3) {$a \succ c \succ b$};
        \node[pref,align=center] (d) at (6,0) {$b \succ c \succ a$};
        
        \path (c) edge [loop above] node {$c(\{b,c\})=c$} (c);
        \path (d) edge node[right] {$c(\{b,c\})=b$} (c);
        \path (a) edge [bend right] node[left] {$c(\{a,b\})=a$} (b);
        \path (b) edge [bend right] node[right] {$c(\{a,b\})=b$} (a);

    \end{tikzpicture}
    \caption{A visual representation of an agent facing deterministic consumption dependence. On the left, we see that the choice of $a$ induces the preference $b \succ c \succ a$ and the choice of $b$ induces the preference $a \succ c \succ b$. In the choice set $\{a,b\}$, this leads to a uniform distribution over $a \succ c \succ b$ and $b \succ c \succ a$. On the right we see that the choice of $b$ and $c$ induce the preference $a \succ c \succ b$. In the choice set $\{b,c\}$, this leads to a distribution that puts full weight on $a \succ c \succ b$.}
    \label{fig:easyexample}
\end{figure}
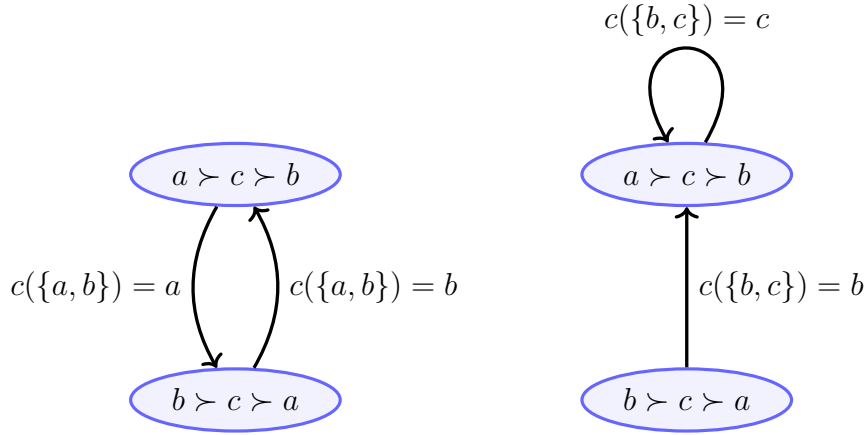

To understand how consumption dependence causes misspecification of the random utility model, consider the following deterministic example. An agent has two possible preferences over apples ($a$), bananas ($b$), and cake ($c$) and is subject to consumption dependence. If the agent just ate an apple, their preference is given by $b \succ c \succ a$. If the agent just ate either a banana or a slice of cake, their preference is given by $a \succ c \succ b$. When our agent repeatedly faces the choice set $\{a, b\}$, eating an apple will induce a preference that causes the agent to eat a banana and eating a banana will induce a preference that causes the agent to eat an apple. This means that our agent will be faced with $a \succ c \succ b$ half of the time and $b \succ c \succ a$ half of the time when their choice set is $\{a, b\}$. Now suppose that our agent's choice set is given by $\{b, c\}$. In this case, eating a banana will induce a preference that causes the agent to eat cake and eating cake will induce a preference that causes the agent to eat cake again. This means that our agent will always face the preference $a \succ c \succ b$. In this example, the distribution over preferences at $\{a, b\}$ and $\{b, c\}$ differ, and thus the random utility model is not an accurate model of this type of behavior. Figure \ref{fig:easyexample} provides a visual representation of this example.

Our first goal in this paper is to understand if the misspecification caused by consumption dependence leads to issues when taking the random utility model to data. In one example, we consider an agent who is subject to persistent cravings. A craving persists only if the agent is unable to sate their craving, so this is consumption dependent behavior. In our example, we show that the introduction of persistence directly causes one of the main axioms of random utility to fail. In a second example, we consider a logit model of habit formation. We show that the addition of habit formation not only causes the standard logit axiom to fail but also causes the standard logit estimator of valuation to be biased cardinally and ordinally.

Our second goal is to characterize exactly when and by how much the random utility model is misspecified in the presence of consumption dependence. Our model allows for both consumption dependence and state dependence. State dependence is the idea that an agent's utility function depends on an underlying state of the world which varies (exogenously) over time. As a preliminary result, we show that, in the absence of state dependence, any amount of consumption dependence leads to misspecification. Once we allow for both consumption and state dependence, we find that there are cases which allow for consumption dependence and are consistent with random utility. Our main theorem shows that these cases are characterized by the distribution over preferences tomorrow when an agent chooses $x$ being equal to the distribution over preferences tomorrow when an agent would choose $x$ but is forced to choose some other alternative. Building on this result, we show that the degree of misspecification is a function of the difference between these two distributions over preferences tomorrow and the mean passage times of the underlying Markov chain over preferences. We also offer a second characterization of when consumption dependence leads to misspecification through a type of no investment condition.

The rest of this paper is organized as follows. In Section \ref{Model} we introduce notation as well as our base model. In Section \ref{Examples}, we present several examples of consumption dependent behavior and show through these examples how consumption dependence leads to misspecification and bias. In Section \ref{Menu Invariance}, we characterize when and by how much consumption dependence leads to misspecification. We conclude and discuss the related literature in Section \ref{Discussion}.

\section{Model}\label{Model}
\subsection{Preliminaries}
Let $X$ be a finite set of alternatives with typical elements $x, y,$ and $z$. We use $\mathcal{X}$ to denote the collection of subsets of $X$ with at least two elements. $\mathcal{L}(X)$ denotes the set of linear orders of $X$ with typical element $\succ$. We let $\Delta(\mathcal{L}(X))$ denote the set of probability distributions over $\mathcal{L}(X)$ with typical element $\nu$. Further, let $int \Delta(\mathcal{L}(X))$ denote the set of full support probability distributions over $\mathcal{L}(X)$. We use $M(\succ,A)$ to denote the element $x \in A$ that maximizes $\succ$ in $A$. Further, we use the shorthand to $x \succ A$ to denote that $x \succ y$ for all $y \in A$ with $A \succ x$ defined analogously. Define $N(x,A)=\{\succ|x \succ A \setminus \{x\}\}$ which denotes the set of linear orders maximized by $x$ in $A$. 

\subsection{Data Generating Process}

Our first goal in this paper is to study the relationship between the random utility model (RUM) and repeated choice when choices are subject to consumption and state dependence. Our notion of data corresponds to the time average of choice and is modeled through a random choice rule.

\begin{definition}
A function $p:X \times \mathcal{X} \rightarrow [0,1]$ is a \textbf{random choice rule} (rcr) if it satisfies the following.
\begin{enumerate}
    \item $p(x,A) \geq 0$
    \item $\sum_{x \in A}p(x,A)=1$
\end{enumerate}
\end{definition}

Our focus, however, is on the underlying dynamics which induce these data. This is in order to see if those dynamics are consistent with the primitives of RUM. Our primitive is what we call a transition function.

\begin{definition}
    A function $t:X \times \mathcal{L}(X) \rightarrow \Delta(\mathcal{L}(X))$ a \textbf{transition function}. Further, we call a function $t:X \times \mathcal{L}(X) \rightarrow int \Delta(\mathcal{L}(X))$ a \textbf{full support transition function}.
\end{definition}

We use the notation $t_{\succ'}(x,\succ)$ to denote the probability put on $\succ'$ when $(x,\succ)$ is the input of $t$. The point of a transition function is to model consumption and state dependence. It is obvious that the consumption input, $x \in X$, models consumption dependence. What is perhaps a bit more subtle is the fact that the preference input, $\succ \in \mathcal{L}(X)$, models state dependence. We use a preference as a sufficient statistic for an underlying state.

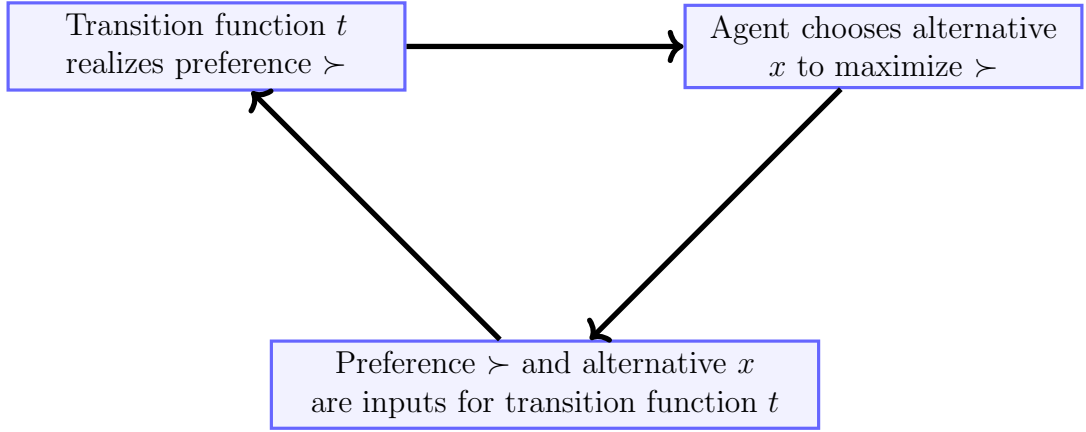
\begin{figure}
    \centering
\begin{tikzpicture}[set/.style={rectangle, draw=blue!60, fill=blue!5, very thick, minimum size=7mm},
choice/.style={diamond, draw=red!60, fill=red!5, very thick, minimum size=7mm},
attention/.style={rectangle, draw=green!60, fill=green!5, very thick, minimum size=7mm}]

        \node[set,text width=5cm, align=center] (h) at (0,0) {Transition function $t$ \\ realizes preference $\succ$};
        \node[set,text width=7cm,align=center] (i) at (4.5,-4.5) {Preference $\succ$ and alternative $x$ \\ are inputs for  transition function $t$};
        \node[set,text width=5cm,align=center] (x) at (9,0) {Agent chooses alternative \\ $x$ to maximize $\succ$};

        \draw [->,line width=2pt] (i) -- (h) node[midway, below, sloped, scale=.8] {};
        \draw [->,line width=2pt] (h) -- (x) node[midway, below, sloped, scale=.8] {};
        \draw [->,line width=2pt] (x) -- (i) node[midway, below, sloped, scale=.8] {};

    \end{tikzpicture}
    \caption{A visual representation of the data generating process. At the start of a period, a preference $\succ$ is realized. The agent then chooses $x$ to maximize $\succ$. Next period's preference is then determined by $t(x,\succ)$.}
    \label{fig:DGP}
\end{figure}

Our model of repeated choice proceeds as follows. In any given period, our agent faces a decision from a fixed choice set $A$. At the start of the period, an agent realizes a preference according to their transition function $t$. Once this preference is realized, the agent chooses the alternative $M(\succ,A)$ in set $A$ which maximizes their realized preference. After making this choice, $M(\succ,A)$ and $\succ$ are used as the inputs to transition function $t$ and $t(M(\succ,A),\succ)$ is used to realize next period's preference. Figure \ref{fig:DGP} offers a visual representation of our data generating process. There are two important things to note about our model. First is that if an agent is choosing from choice set $A$ in some period, then we assume they are always choosing from choice set $A$.\footnote{We consider an extension in in the appendix where the agent's choice set is allowed to vary over time.} Because of this, we are able to interpret $p(x,A)$ as the time average choice of an agent who always faces choice set $A$. The second thing to note is that our agent is myopic. While the realization of their preference depends on their history of choices and preferences, we assume that our agent does not take into account how their choice today impacts their realized preference tomorrow. While this is not a model of a hyper-rational agent, we find this assumption to be a reasonable approximation of high frequency decisions, such as cereal or beer choice, where consumers may neglect the impact of any one choice on their future choices. This also allows us to isolate the impact of consumption dependence in the intertemporal aggregation problem.\footnote{See \citet{lu2020repeated} for a discussion of how forward looking behavior impacts the intertemporal aggregation problem.}

Given that an agent's choice set is fixed over time, we make the observation that a fixed choice set and a fixed transition function $t$ induce a Markov chain over preferences. Once an agent realizes a preference $\succ$ and is faced with choice set $A$, their decision is pinned down to be $M(\succ,A)$, the maximal element of $A$ according to $\succ$. This pins down the inputs to transition function $t$ as $M(\succ,A)$ and $\succ$. Following this logic, for every choice set $A$, we can think of the Markov chain over preference defined by the following matrix.
\begin{equation}
    M_A=\begin{bmatrix} m_A(\succ,\succ')=t_{\succ'}(M(\succ,A),\succ) \end{bmatrix}
\end{equation}
The rows and columns of $M_A$ are indexed by the elements of $\mathcal{L}(X)$. The element $m_A(\succ,\succ')$ denotes the element of $M_A$ in row $\succ$ and column $\succ'$ and encodes the transition probability from preference $\succ$ to preference $\succ'$ for choice set $A$. When $t$ is a full support transition function, the Markov chain at choice set $A$ is ergodic and thus has a unique stationary distribution.\footnote{For the results in Section \ref{Menu Invariance}, we assume a full support transition function. However, all theorems in this section can be extended to general transition functions and their \textit{invariant} distributions. We consider these extensions in the appendix.} We use $\nu_A$ to denote the unique stationary distribution of the Markov chain over preferences at choice set $A$.

In our model, $\nu_A$ corresponds to the time average of preferences faced by our agent at choice set $A$. Accordingly, we assume that the time average of choices should correspond to the time average of preferences, and thus our model is dictated by the following equation.
\begin{equation}\label{MRUM}
    p(x,A)=\sum_{\succ \in N(x,A)}\nu_A(\succ)
\end{equation}
Notably, our model is an extension of RUM as proposed in \citet{block1959random}. In classic RUM, the distribution over preferences is the same at every choice set. However, in our model the distribution over preferences is allowed to vary between choice sets according to the underlying transition function. Our main goal in Section \ref{Menu Invariance} is characterizing when these two models coincide.

\section{Examples}\label{Examples}

In the previous section, we introduced transition functions as our main tool for modeling consumption and state dependence. Our goal in this section is twofold. First, through a series of examples, we show that transition functions are a flexible tool for modeling various consumption and state dependent behaviors. Our second goal is to motivate our study of the intertemporal aggregation problem in the presence of consumption dependence. Through two examples, we show that, in the presence of certain types of consumption dependence, standard axioms of random utility fail and standard discrete choice estimators are biased. In one of these examples, we argue that these failures are due to the fact that the underlying distribution over preferences, $\nu_A$, necessarily varies with the choice set. We now offer four examples of consumption and state dependent behavior as well as assumptions put on transition functions which model these behaviors.

\begin{example}[Learning Through Experience]
    Consider an agent who has a prior belief about the utility of each of a set of alternatives. This agent can learn about an alternative's utility only by consuming that good. This type of learning implies that when an agent consumes alternative $x$, their ranking of $y$ and $z$ remains the same. This can be modeled using transition functions in the following way.
    \begin{equation*}
        t_{\succ'}(x,\succ)\begin{cases}
                >0 \text{ if } y \succ' z \text{ when } y \succ z \text{ for } x\neq y,x\neq z \\
                =0 \text { otherwise}
            \end{cases}
    \end{equation*}
\end{example}

\begin{example}[Habit Formation and Preference for Variety]
    Consider two agents. The first agent is subject to habit formation and receives higher utility from an alternative if they consumed that alternative yesterday. The second agent has a preference for variety and receives lower utility from an alternative if they consumed that alternative yesterday. We can model habit formation as the following.
    \begin{equation*}
                t_{\succ'}(x,\succ)\begin{cases}
                >0 \text{ if } x \succ' y \text{ when } x \succ y \text{ and } y \succ' z \text{ when } y \succ z  \text{ for } x\neq y,x\neq z \\
                =0 \text { otherwise}
            \end{cases}
    \end{equation*}
    Similarly, we can model preference for variety as follows.
    \begin{equation*}
                t_{\succ'}(x,\succ)\begin{cases}
                >0 \text{ if } y \succ' x \text{ when } y \succ x \text{ and } y \succ' z \text{ when } y \succ z  \text{ for } x\neq y,x\neq z \\
                =0 \text { otherwise}
            \end{cases}
    \end{equation*}
    While we use Markovian transition functions, we note that the transition functions in this example implicitly allow for dependence on longer histories of choice. In both cases, the preference input acts as an endogenous state. Repeated consumption of $x$ will further raise/lower the ranking of $x$ in the agent's preference in the case of habit formation/preference for variety.
\end{example}

\begin{example}[Stochastic Reference Dependence with Status Quo Bias]
    Imagine an agent who is reference dependent in that the realization of their preference today depends on the reference point of the agent. In this example, the agent's reference point is exactly the alternative they consumed yesterday. The agent's preference is allowed to be random, but status quo bias imposes that $x$ is more likely to be preferred to $y$ when $x$ is the reference point than when $z$ is the reference point. The following restriction on a transition function captures this behavior.
    \begin{equation*}
    \begin{split}
        \text{For all } (\succ,\succ') \text{ and } x, \text{ } t(x,\succ)&=t(x,\succ') \\
        \text{For all } \succ,\text{ }  \sum_{\succ'} t_{\succ'}(x,\succ) \mathbf{1}\{x \succ' y\} &\geq \sum_{\succ'} t_{\succ'}(z,\succ) \mathbf{1}\{x \succ' y\}
    \end{split}
    \end{equation*}
\end{example}

\begin{example}[Intertemporal Complements and Substitutes]
    Consider an agent who chooses between alternatives, some of which are complements or substitutes for other alternatives. If $x$ and $y$ are complements, then choosing $x$ today raises the utility of $y$ tomorrow. Similarly, if $x$ and $y$ are substitutes, then choosing $x$ today lowers the utility of $y$ tomorrow. Suppose $x$ is only a complement of $y$, then we can model this behavior as follows.
    \begin{equation*}
        t_{\succ'}(x,\succ)\begin{cases}
                >0 \text{ if } y \succ' z \text{ when } y \succ z\\
                =0 \text { otherwise}
            \end{cases}
    \end{equation*}
    Similarly, suppose $x$ is only a substitute of $y$. We can then model this as follows.
        \begin{equation*}
        t_{\succ'}(x,\succ)\begin{cases}
                >0 \text{ if } z \succ' y \text{ when } z \succ y\\
                =0 \text { otherwise}
            \end{cases}
    \end{equation*}
    Notably, this representation of complements and substitutes allows for one directional complementarity and substitutability. Without further imposing it as the analyst, the above assumptions do not require that $x$ being a complement/substitute for $y$ implies that $y$ is a complement/substitute for $x$.
\end{example}

\subsection{Persistent Cravings}\label{PersistenCravingsRCR}

We now turn our attention to the random cravings model of \citet{honda2021random}. In this model, an agent has a base preference $\rhd$ but is subject to cravings. When an agent has a craving for alternative $x$, they face preference $\succ_x$ which is the same as $\rhd$ except that $x$ is ranked highest. In \citet{honda2021random}, the distribution over cravings (i.e. the distribution over preferences) is fixed between choice sets. We build on this model by making the assumption that cravings can persist if they are not sated. Once persistence is added to the model, it turns out that this extended model fails to satisfy a standard random utility axiom.

\begin{definition}
    A set of preferences $\{\succ_x\}_{x \in X}$ satisfies the \textbf{random cravings property} with respect to linear order $\rhd$ if the following conditions hold.
    \begin{enumerate}
        \item $\forall x \in X$, $x \succ_x X \setminus \{x\}$
        \item $\forall x,y,z \in X$ with $y\neq x$ and $z\neq x$, $y \succ_x z$ if and only if $y \rhd z$
    \end{enumerate}
\end{definition}

In the above definition, $\rhd$ can be thought of the agent's base preference and each preference $\succ_x$ corresponds to when the agent craves $x$.

\begin{definition}
    A distribution over preferences $\nu$ whose support satisfies the random cravings property with respect to $\rhd$ is \textbf{craving monotonic} if $x \rhd y$ implies $\nu(\succ_x)>\nu(\succ_y)>0$.
\end{definition}

In a craving monotonic representation, every alternative is craved with positive probability and the probability that each alternative is craved is ranked by $\rhd$. This assumption is simply asking that more preferred alternatives are craved more often. Thus far, the setup of our model has been exactly the same as in \citet{honda2021random}. We now introduce persistence functions and extend the model of \citet{honda2021random}.

\begin{definition}
    A function $\phi:X^2 \rightarrow [0,1)$ is a \textbf{persistence function} if it satisfies the following.
    \begin{enumerate}
        \item $\phi(x,x)=0$
        \item $\phi(x,y)>0$ for all $x \neq y$
    \end{enumerate}
\end{definition}

We use persistence functions to model the persistence of cravings. The probability that a craving for $y$ persists when the agent consumes $x$ is given by $\phi(x,y)$. The above definition then tells us that cravings never persist when sated (i.e. the agent consumes $x$ when they crave $x$), and persist with positive probability when not sated (i.e. the agent consumes $x$ when they crave $y$). We now define the persistent craving model.

\begin{definition}
    A random choice rule $p$ is \textbf{consistent} with the persistent craving model if there exists a persistence function $\phi$, a linear order $\rhd$, a distribution over preferences $\nu$ whose support satisfies the random cravings property and is craving monotonic with respect to $\rhd$, and a transition function $t$ such that the following two equations hold for all $A \in \mathcal{X}$, $x \in A$, and $(x,\succ) \in X \times \mathcal{L}(X)$.
    \begin{equation}\label{PCMChoiceProbs}
        p(x,A)=\sum_{\succ \in N(x,A)}\nu_A(\succ)
    \end{equation}
    \begin{equation}\label{PCMTransitionFunctions}
        t(x,\succ_y)=\phi(x,y) \delta_{\succ_y} + (1-\phi(x,y))\nu
    \end{equation}
    Further, we suppose that $t(x,\succ)=\nu$ for each $\succ$ not in the support of $\nu$. We say that such a transition function $t$ has a persistent craving representation.
\end{definition}

In the prior definition, we use $\delta_{\succ_y}$ to denote the degenerate distribution that puts full weight on linear order $\succ_y$. Equation \ref{PCMChoiceProbs} tells us that the persistent craving model is in line with our general model from Section \ref{Model}. Equation \ref{PCMTransitionFunctions} then tells us that the transition function of the model imposes persistent cravings. Our main question in regards to the persistent craving model is whether or not it is a random utility model. If it is, then the persistent craving model must satisfy the axioms of random utility. The main normative axiom of random utility is regularity.

\begin{definition}
    A random choice rule $p$ satisfies \textbf{regularity} if $x \in A \subseteq B$ implies that $p(x,A) \geq p(x,B)$.
\end{definition}

Regularity simply asks that the choice probability of each alternative increases as we move down in the set inclusion order. Regularity can be thought of a stochastic version of Sen's $\alpha$, and it is a necessary axiom of random utility as each preference that chooses $x$ in $B$ still chooses $x$ in $A$.

\begin{proposition}\label{PCMReg}
    Suppose that $|X|\geq 3$. If a random choice rule $p$ is consistent with the persistent craving model, then it does not satisfy regularity.
\end{proposition}

As Proposition \ref{PCMReg} shows, the persistent craving model is not a random utility model as it fails to satisfy regularity. We can actually connect the size of regularity failures directly with our persistence function.

\begin{proposition}\label{PCMComparative}
    Consider two persistent craving representations $(\rhd,\nu,\phi)$ and $(\rhd,\nu,\phi')$ with associated random choice rules $p$ and $p'$. For $y \in X \setminus \{x\}$ with $y \neq M(\rhd,X \setminus \{x\})$, $p(y,X\setminus \{x\}) - p(y,X) > p'(y,X\setminus \{x\}) - p'(y,X)$ if and only if $\phi(M(\rhd,X\setminus \{x\}),x) > \phi'(M(\rhd,X\setminus \{x\}),x)$.
\end{proposition}

Proposition \ref{PCMComparative} tells us that the persistent craving model's degree of departure from random utility is directly and monotonically related to the size of the persistence parameter in the model. This means that stronger consumption dependence leads to larger failures of regularity. Since the persistent craving model is contained by the model described in Section \ref{Model}, we know that there exists some distribution over preferences governing choice at each choice set. However, because the persistent craving model is not a random utility model, we know that it is necessarily the case that this distribution over preferences varies with the choice set.

\begin{proposition}\label{PCMMI}
    A transition function $t$ with a persistent cravings representation satisfies $\nu_X \neq \nu_{X \setminus \{x\}}$.
\end{proposition}

Together, Propositions \ref{PCMReg}- \ref{PCMMI} tell us that our transition functions truly lead to menu dependence of the underlying distribution over preferences. Further, they tell us that this menu dependence actually leads to problems when we want to use standard random utility tools and axioms. This motivates our discussion in Section \ref{Menu Invariance} where our aim is to characterize exactly which transition functions lead to menu invariant distributions over preferences.

\subsection{Logit Habit Formation}\label{HabitFormationRCR}

Perhaps the most used random utility model in applied settings is the Luce/logit model and its variants. Just as the logit model is a special case of the random utility model, there is an analogue of the logit model in our setting which is a special case of our general setup.\footnote{See \citet{block1959random} for a proof of the classic result on Luce and RUM. The fact that our extension of logit is a subset of our model follows from an analogous proof.} In this section, we focus on an extension of logit which allows for Markovian habit formation. We model this by asking that the utility of good $x$ is higher when good $x$ was consumed yesterday. While transition functions can accommodate this behavior, in this example, we keep with the norm of working in cardinal space when working with the logit model. This allows us to discuss parameter estimation As such, we use the following to describe the conditional choice probabilities of our logit model.

\begin{equation}\label{LogitCCP}
    p(x,A|y)=\frac{e^{v(x)+c(x) \mathbf{1}\{x=y\}}}{\sum_{z \in A} e^{v(z)+c(z) \mathbf{1}\{z=y\}}}
\end{equation}

Just as was the case with transition functions, these conditional choice probabilities define a Markov chain at each menu. However, the major difference is that transition functions define a Markov chain over preferences and these conditional choice probabilities define a Markov chain over alternatives. Given the nature of the logit model and Equation \ref{LogitCCP}, each of these Markov chains are ergodic and thus have a unique stationary distribution. This leads us to our representation.

\begin{definition}\label{LogitHabitFormation}
    A random choice rule $p$ is \textbf{consistent} with habit formation logit if there exists functions $v:X \rightarrow \mathbb{R}$ and $c:X \rightarrow \mathbb{R}^+$ such that, for all sets $A \in \mathcal{X}$, $p(\cdot,A)$ is equal to the stationary distribution of the Markov chain defined by Equation \ref{LogitCCP}.
\end{definition}

Further, keeping with applied work that uses the logit model, we assume that there is some outside option $o$ which satisfies $v(o)=0$, $c(o)=0$, and is available in every choice set.\footnote{Note that $v(o)=0$ is the standard assumption and is simply a normalization. The assumption that $c(o)=0$ is not just a normalization assumption and has behavioral content. Nonetheless, we maintain it for our exposition.} Just as we discussed with the persistent craving model, we now ask if habit formation logit is contained within the standard logit model. Logit is characterized by two axioms; positive choice probabilities and independence of irrelevant alternatives \citep{luce1959individual}.

\begin{axiom}[IIA]\label{IIA}
    A random choice rule $p$ satisfies \textbf{independence of irrelevant alternatives} if for all $x,y \in A \cap B$ we have that $\frac{p(x,A)}{p(y,A)}=\frac{p(x,B)}{p(y,B)}$.
\end{axiom}

Since the conditional choice probabilities in habit formation logit put positive weight on each alternative, we know that habit formation logit satisfies positivity.

\begin{proposition}\label{failIIA}
    A random choice rule $p$ with a habit formation logit representation satisfies IIA if and only if $c(x)=0$ for all $x \in X$.
\end{proposition}

Proposition \ref{failIIA} tells us that habit formation logit is in fact not contained by the standard logit model. The intuition behind this result becomes apparent when looking at the closed form representation of $p(x,A)$.
\begin{equation}\label{HabitLogitStat}
     p(x,A)=\frac{e^{v(x)}\left(\sum_{y\in A} e^{v(y) + c(y)\mathbf{1}\{x=y\}}\right)}{\sum_{z \in A}e^{v(z)}\left(\sum_{y\in A} e^{v(y) + c(y)\mathbf{1}\{z=y\}}\right)}
\end{equation}

Equation \ref{HabitLogitStat} shows that habit formation logit choice probabilities take a similar form to standard logit choice probabilities. The notable difference between the two is the summation which follows the standard logit term $e^{v(x)}$. The reason that IIA fails is because the term $\sum_{y\in A} e^{v(y) + c(y)\mathbf{1}\{x=y\}}$ is both alternative and choice set dependent.

As mentioned at the start of this example, our focus on the logit case with an outside option is to consider parameter estimation. When working with logit with an outside option, the utilities of each alternative are identified and the standard estimator for these utilities are given by the following.

\begin{equation}\label{logitIdentification}
    \hat{v}(x)=\log\left(\frac{p(x,\{x,o\})}{p(o,\{x,o\})}\right)
\end{equation}

\begin{proposition}\label{biasedEstimator}
    In the habit formation logit model, $\hat{v}(x)$ is unbiased if and only if $c(x)=0$.
\end{proposition}

Just as in Proposition \ref{failIIA}, Proposition \ref{biasedEstimator} tells us that the standard logit tools only work when we have no habit formation and thus no consumption dependence. To understand how the bias arises, the following tells us what $\hat{v}(x)$ actually estimates in the habit formation logit model.

\begin{equation}\label{logitbiaseq}
    \hat{v}(x)= v(x) + \log(1+e^{v(x)+c(x)}) - \log(1+e^{v(x)})
\end{equation}

In the standard logit setup, the last two terms of Equation \ref{logitbiaseq} do not appear. Equation \ref{logitbiaseq} actually tells us more than just Proposition \ref{biasedEstimator}. For sufficiently strong habit formation (high values of $c(x)$), it is possible for alternatives with $v(x)<0$ to have estimated valuations $\hat{v}(x)>0$.\footnote{If we were to consider an analogous model of preference for variation in which $c(x) \leq 0$, we would also get the opposite direction. That is, when $c(x)$ is allowed to be negative, the standard estimator may estimate a negative utility value when the actual utility value is positive.} Now suppose that we have more than two alternatives, not counting the outside option, and we wish to rank alternatives by their utility level, $v(x)$. If alternatives have sufficiently similar valuations but sufficiently different $c$ terms, the ranking of the estimated values $\hat{v}(x)$ will differ from the true ranking of $v(x)$. This is all to say that if we ignore consumption dependence when we aggregate across time, not only will we face biased estimators, but our estimators may even be incorrect in terms of their ordinal and directional implications.

\section{Menu Invariance}\label{Menu Invariance}

In the last section, we saw how ignoring consumption dependence when we aggregate across time can cause many of the standard random utility tools to fail. Our main goal in this section is to characterize which forms of consumption dependence are consistent with the classic model of random utility. To do this, we study transition functions and ask exactly which transition functions lead to a menu invariant distribution over preferences. In this section we maintain the assumption that each transition function $t$ is a full support transition function.\footnote{We relax this assumption for our main characterizations in the appendix.}

\begin{definition}\label{MenuInvDef}
    A full support transition function $t$ is \textbf{menu invariant} if $\nu_A=\nu_B$ for all sets $A,B \in \mathcal{X}$.
\end{definition}

In simple terms, a transition function $t$ is menu invariant if it has a random utility representation. Before we move on to the general case, we first focus on the special cases of consumption independence and state independence. These two cases are the two extremes of our model.

\begin{definition}
    A transition function $t$ is \textbf{consumption independent} if, for all $x,y \in X$ and for all $\succ \in \mathcal{L}(X)$, we have $t(x,\succ)=t(y,\succ)$. In the case of consumption independence, we write $t(\succ)$ instead of $t(x,\succ)$.
\end{definition}

Recall that each choice set $A$ has a Markov chain over preferences that is dictated by $t(M(\succ,A),\succ)$. The only way that these Markov chains differ from set to set is through the consumption input term in the transition function. However, in the case of consumption independent transition functions, the consumption input never actually impacts the underlying Markov chains. It immediately follows from this observation that every consumption independent transition function is menu invariant.

\begin{observation}\label{CIMenuInv}
    If a full support transition function $t$ is consumption independent, then it is menu invariant.
\end{observation}

Observation \ref{CIMenuInv} tells us that, if consumption dependence is not present, then we are free to use any random utility tool we please to analyze our model. Now consider the other extreme case of our model.

\begin{definition}
    A transition function $t$ is \textbf{state independent} if, for all $x \in X$ and for all $\succ, \succ' \in \mathcal{L}(X)$, we have $t(x,\succ)=t(x,\succ')$. In the case of state independence, we write $t(x)$ instead of $t(x,\succ)$.
\end{definition}

State independent transition functions are the exact opposite of consumption independent transition functions. We just mentioned that the Markov chains over preferences associated with each choice set only differ through the consumption input term of our transition function. In the case of state independent transition functions, the consumption input is the only input that matters in our transition function. This turns out to mean that, once we have any meaningful amount of consumption dependence, any state independent transition function fails to be menu invariant.

\begin{proposition}\label{SIMenuInv}
    Suppose that $|X|\geq 3$. If full support transition function $t$ is state independent, then $t$ is menu invariant if and only if $t(x)=t(y)$ for all $x,y \in X$.
\end{proposition}

Among our examples in Section \ref{Examples}, both habit formation logit and stochastic reference dependence with status quo bias are state independent. As such, once we have meaningful habit formation or meaningful reference dependence, both of these models fail to be consistent with random utility. This provides another explanation for our results on the failure of logit axioms and estimators in habit formation logit. Proposition \ref{SIMenuInv} leaves us with the question of if there exists any transition function which allows for meaningful consumption dependence while being menu invariant. Example \ref{MeaningfulConDep} offers one such transition funciton.

\begin{example}\label{MeaningfulConDep}
    Let $X = \{x,y,z\}$. To keep this example simple, we restrict our attention to two preferences. Let $\succ_x$ rank $x \succ_x y \succ_x z$ and $\succ_z$ rank $z \succ_z y \succ_x$. Our transition function is described as follows.
    \begin{equation}\label{exampletransitionfunciton}
        \begin{split}
            t(x,\succ) = \begin{cases}
                \frac{2}{3} \text{ } \succ_x \\
                \frac{1}{3} \text{ } \succ_z
            \end{cases} & t(z,\succ) =\begin{cases}
                \frac{2}{3} \text{ } \succ_z \\
                \frac{1}{3} \text{ } \succ_x
            \end{cases} \\
            t(y,\succ_x) = \begin{cases}
                \frac{2}{3} \text{ } \succ_x \\
                \frac{1}{3} \text{ } \succ_z
            \end{cases} & t(y,\succ_z) = \begin{cases}
                \frac{2}{3} \text{ } \succ_z \\
                \frac{1}{3} \text{ } \succ_x
            \end{cases}
        \end{split}
    \end{equation}
    When our transition function is described by Equation \ref{exampletransitionfunciton}, the Markov chain over preferences at each non-singleton subset of $X$ is given by the following.
        \begin{equation}
    M_A=\begin{blockarray}{ccc}
 & \succ_x & \succ_z \\
\begin{block}{c[cc]}
  \succ_x & 2/3 & 1/3 \\
  \succ_z & 1/3 & 2/3 \\
\end{block}
\end{blockarray}
\end{equation}
As $t(y,\cdot)$ depends on the preference input, we have meaningful state dependence. Since $t(x,\succ)$ and $t(z,\succ)$ differ, we have meaningful consumption dependence. Further, since $M_A$ does not depend on $A$, it then follows that $\nu_A$ does not depend on $A$ and thus we have menu invariance.
\end{example}

The main take away from Example \ref{MeaningfulConDep} is that, when $y$ is chosen but $x$ is the most preferred alternative, $t(y,\cdot)$ acts as if $x$ was chosen. Similarly, when $y$ is chosen but $z$ is the most preferred alternative, $t(y,\cdot)$ acts as if $z$ was chosen. While this is stronger than what we need to characterize menu invariance, as we will see in the next section, menu invariant transition functions are characterized by the transition function behaving on average as if the most preferred alternative was chosen when the second most preferred alternative is chosen.

\subsection{Local Invariance}\label{Local Invariance}

We now focus on characterizing menu invariance for general transition functions. In this section, we focus on our first of two characterizations. This first characterization builds on the intuition of Example \ref{MeaningfulConDep} and asks that transition functions on average act as if the most preferred alternative is chosen when the second most preferred alternative is chosen instead. We call this condition local invariance.

\begin{definition}\label{LocalInvDef}
    We say that a transition function $t$ is \textbf{locally invariant} with respect to distribution $\nu$ if, for all $x \in A \in \mathcal{X}$ with $|A|\geq 3$, we have the following.
        \begin{equation}\label{LocInv}
        \underbrace{\sum_{\succ \in N(x,A)} \nu(\succ)t(x,\succ)}_{\substack{\text{Distribution over preferences}\\ \text{after agent chooses $x$ from $A$}}} = \underbrace{\sum_{y \in A \setminus \{x\}} \sum_{\succ \in N(x,A) \cap N(y,A \setminus \{x\})}  \nu(\succ)t(y,\succ)}_{\substack{\text{Distribution over preferences after agent would} \\ \text{choose $x$ from $A$ but has to choose from $A \setminus \{x\}$}}}
    \end{equation}
\end{definition}

Let us think about Equation \ref{LocInv} in relation to $A$ and $A \setminus \{x\}$. Suppose the agent draws a preference $\succ$ which chooses some alternative $y \neq x$ from $A$. Since $y \in A \setminus \{x\}$, it follows that $\succ$ will also choose $y$ from $A \setminus \{x\}$. This means that no matter what distribution over preferences $A$ and $A \setminus \{x\}$ face today, the distribution over preferences tomorrow, conditional on drawing a preference today which chooses any alternative $y \neq x$, is the same at sets $A$ and $A \setminus \{x\}$. This tells us that when we compare the unconditional distribution over preferences tomorrow, we only need to worry about the behavior of our transition function when we draw preferences which choose $x$ from $A$. Equation \ref{LocInv} is exactly the condition that guarantees the unconditional distribution over preferences tomorrow is the same at $A$ and $A\setminus \{x\}$.

\begin{theorem}\label{LocalInvThm}
    Given a full support transition function $t$, the following are equivalent.
    \begin{enumerate}
        \item $t$ is menu invariant.
        \item $t$ is locally invariant with respect to $\nu_A$ for all $A\in \mathcal{X}$.
        \item $t$ is locally invariant with respect to $\nu_A$ for any $A \in \mathcal{X}$.
    \end{enumerate}
\end{theorem}

Theorem \ref{LocalInvThm} shows that local invariance with respect to any stationary distribution $\nu_A$ characterizes menu invariance. The point of this characterization is to offer a simple test through local invariance which can be applied to a model before it is brought to data. The emphasis here is on simple. In theory, one could directly calculate the stationary distribution over preferences at each choice set to directly test menu invariance given a transition function. Local invariance allows an analyst to simply ask whether the behavior after choosing $x$ can possibly be the same as the behavior after choosing $y$ when an agent wants to choose $x$. While local invariance makes direct reference to some underlying stationary distribution, it is often possible to argue that a transition function fails local invariance without reference to any distribution. As we show in Section \ref{LocInvExamples}, for every example we consider in the paper which is not already state independent, we can argue that they fail local invariance without reference to an explicit stationary distribution. A second benefit of this characterization is that there is a deep relationship between the degrees by which local invariance and menu invariance fail. As we will see in Section \ref{Failures of LI}, the degree by which menu invariance fails depends only on the degree by which local invariance fails and the mean passage time of the underlying Markov chains.

\subsubsection{Local Invariance Through Examples}\label{LocInvExamples}

In this section, we apply our Theorem \ref{LocalInvThm} to the remaining examples of Section \ref{Examples}. We show that all of these examples fail local invariance and thus fail menu invariance. This tells us that we are unable to use random utility tools in the presence of each of these behaviors.

\begin{example}[Learning Through Experience Revisited]
    Recall that learning through experience is modeled through the following restriction on transition functions.
    \begin{equation*}
        t_{\succ'}(x,\succ)\begin{cases}
                >0 \text{ if } y \succ' z \text{ when } y \succ z \text{ for } x\neq y,x\neq z \\
                =0 \text { otherwise}
            \end{cases}
    \end{equation*}
    Suppose that we are in the case where each alternative's true utility is higher than the agent's expectation of that alternative's utility. This means that when $x$ is chosen today, the relative ranking of $x$ can only (weakly) increase. However, when the agent would prefer to choose $x$ but is forced to choose $y$ instead, the relative ranking of $x$ can only (weakly) decrease. As such, there are cases when learning through experience fails local invariance and thus fails to be menu invariant.
\end{example}

\begin{example}[Habit Formation and Preference for Variety Revisited]
    Recall that habit formation is modeled through the following restriction on transition functions.
        \begin{equation*}
                t_{\succ'}(x,\succ)\begin{cases}
                >0 \text{ if } x \succ' y \text{ when } x \succ y \text{ and } y \succ' z \text{ when } y \succ z  \text{ for } x\neq y,x\neq z \\
                =0 \text { otherwise}
            \end{cases}
    \end{equation*}
    Suppose that $x$ is preferred to $y$ today and our agent chooses $x$. The agent's realized preference tomorrow must satisfy $x \succ' y$. However, suppose that $x$ is unavailable and the agent chooses $y$ instead. Given sufficiently strong habit formation, it is possible that tomorrow's realized preference satisfies $y \succ' x$. This is a failure of local invariance and thus a failure of menu invariance. An analogous argument can be made for preference for variation just with the ranking reversed for tomorrow's preference.
\end{example}

\begin{example}[Intertemporal Complements and Substitutes Revisited]
    Recall that intertemporal complementarity is modeled through the following restriction on transition functions.
        \begin{equation*}
        t_{\succ'}(x,\succ)\begin{cases}
                >0 \text{ if } y \succ' z \text{ when } y \succ z\\
                =0 \text { otherwise}
            \end{cases}
    \end{equation*}
    Suppose that $x$ is the only complement of $y$. When $x$ is chosen today, the relative ranking of $y$ can only improve tomorrow. However, when any other alternative is chosen, the relative ranking of $y$ is allowed to decrease tomorrow. This is a failure of local invariance and thus a failure of menu invariance. For intertemporal substitutes, you can once again get the same result just arguing that the relative ranking of $y$ must decrease tomorrow when $x$ is chosen today.
\end{example}

\begin{example}[Persistent Cravings Revisited]
    Recall that the persistent craving model puts the following restriction on transition functions where $\nu$ is a craving monotonic distribution.
        \begin{equation*}
        t(x,\succ_y)=\phi(x,y) \delta_{\succ_y} + (1-\phi(x,y))\nu
    \end{equation*}
    Further recall that $\phi(x,y)=0$ if and only if $x=y$. When $x$ is most preferred and chosen, tomorrow's distribution over preferences is given by $\nu$. However, when $x$ is most preferred but $y$ is chosen, tomorrow's distribution over preferences is a strict convex combination of $\nu$ and the degenerate distribution over $\succ_x$. This is a failure of local invariance and thus a failure of menu invariance.
\end{example}

\subsubsection{Failures of Local Invariance}\label{Failures of LI}

Thus far we have shown that local invariance characterizes menu invariance and that there are many reasonable behaviors which fail to satisfy local invariance. In this section, we study the connection between the size of a local invariance failure and the size of a menu invariance failure. Specifically, when we compare two sets $A$ and $A\setminus \{x\}$ and see that Equation \ref{LocInv} fails to hold, we wish to know how large the difference is between $\nu_A$ and $\nu_{A\setminus \{x\}}$. For the entirety of this section, we restrict our analysis to comparisons between $A$ and $A\setminus \{x\}$ for distributions $\nu_A$ and $\nu_{A\setminus \{x\}}$. As such, define $\epsilon_A=\nu_A[M_A-M_{A \setminus \{x\}}]$ and $\epsilon_{A \setminus \{x\}} =\nu_{A\setminus\{x\}}[M_A-M_{A \setminus \{x\}}]$. $\epsilon_A$ exactly captures the left side of Equation \ref{LocInv} minus the right hand side of Equation \ref{LocInv} when using $\nu_A$ as our distribution. As such, $\epsilon_A$ is able to capture by how much local invariance fails. For our first result, we use what is called the Moore-Penrose inverse of a matrix.

\begin{definition}
    Given a matrix $M$, the \textbf{Moore-Penrose inverse} of $M$ is any matrix $M^{mp}$ satisfying the following.
    \begin{enumerate}
        \item $M M^{mp} M = M$
        \item $M^{mp} M M^{mp} = M^g$
        \item $(M M^{mp})^T = M M^{mp}$
        \item $(M^{mp} M)^T = M^{mp} M$
    \end{enumerate}
\end{definition}
Above, we use $M^T$ to denote the transpose of matrix $M$. The Moore-Penrose inverse is also known as the pseudoinverse of a matrix. The Moore-Penrose inverse of a matrix extends the idea of the inverse of a square matrix to general $m$ by $n$ matrices. The important property that we take advantage of is the fact that, when $M$ has full rank, $M M^{mp}$ is the identity matrix.

\begin{proposition}\label{2eLocFail}
    For a full support transition function $t$, the following captures the relationship between $\epsilon_A$, $\epsilon_{A\setminus\{x\}}$, $\nu_A$, and $\nu_{A \setminus \{x\}}$.
    \begin{equation}\label{2elocfaileq}
        \epsilon_A-\epsilon_{A\setminus\{x\}}=(\nu_A-\nu_{A\setminus\{x\}})[M_A-M_{A \setminus \{x\}}]
    \end{equation}
    Further, if $[M_A-M_{A \setminus \{x\}}]$ has full rank, then the following also holds.
    \begin{equation}\label{2elocfailGenInv}
        (\epsilon_A-\epsilon_{A\setminus\{x\}})[M_A-M_{A \setminus \{x\}}]^{mp}=\nu_A-\nu_{A\setminus\{x\}}
    \end{equation}
\end{proposition}

Proposition \ref{2eLocFail} follows directly from the definition of $\epsilon_A$ and $\epsilon_{A \setminus \{x\}}$. Our next result delves deeper into the relationship between menu and local invariance. For this result, the concept of mean passage time is important. Given an irreducible Markov chain $M$, the mean passage time from state $i$ to state $j$ is the average number of periods it takes to go from state $i$ to state $j$. In our setup, we can think of a matrix $N_A$ where entry $n_A(\succ,\succ')$ encodes the mean passage time from preference $\succ$ to preference $\succ'$ given ergodic Markov chain $M_A$.\footnote{For a closed form expression of such a matrix, see equation (2.14) from \citet{hunter2005stationary}.} For a square matrix $M$, let $M^d$ denote the matrix whose entries agree with $M$ on the diagonal and are zero everywhere else. The first part of the following proposition follows directly from Theorem 2.3 of \citet{hunter2005stationary}.

\begin{proposition}\label{GenFailLocInv}
    For a full support transition function $t$, the following captures the relationship between $\epsilon_{A}$, $\nu_A$, and $\nu_{A \setminus \{x\}}$.
    \begin{equation}\label{GenFailLocInvEq}
        \nu_A-\nu_{A\setminus\{x\}}=\epsilon_{A}(N_{A\setminus\{x\}}^d-N_{A\setminus\{x\}})(N_{A\setminus\{x\}}^d)^{-1}
    \end{equation}
    Further, when we restrict attention to $\nu_{A\setminus \{x\}}(\succ)$, we get the following.
        \begin{equation}\label{FailLocInvSingleEq}
        \nu_{A\setminus \{x\}}(\succ)=\frac{\nu_A(\succ)}{1 - \sum_{\succ' \neq \succ}\epsilon_{A}(\succ')n_{A\setminus \{x\}}(\succ',\succ)}
    \end{equation}
\end{proposition}

Proposition \ref{GenFailLocInv} shows that there is a tight connection between the size of a local invariance failure, $\epsilon_A$, and the size of a menu invariance failure, $\nu_A - \nu_{A \setminus \{x\}}$. To summarize Proposition \ref{GenFailLocInv}, a failure of local invariance $\epsilon_A(\succ)$ tells us the difference in probabilities tomorrow when leaving the state $\succ$. This difference in probability impacts our stationary distribution only through how much it impacts the average return time to $\succ$. Turning our attention to Equation \ref{FailLocInvSingleEq}, we can see that $\epsilon_A(\succ')$ acts as a weight on the mean passage time from $\succ'$ to $\succ$. Given these weights, Equation \ref{FailLocInvSingleEq} tells us that, if the weighted sum of mean passage times is positive (negative), then $\nu_{A \setminus \{x\}}(\succ)$ is larger (smaller) than $\nu_A(\succ)$. We view Proposition \ref{GenFailLocInv}, especially Equation \ref{FailLocInvSingleEq}, as a general tool for studying the intertemporal aggregation problem when menu invariance fails. As an example of this, we take Proposition \ref{GenFailLocInv} to the persistent craving model.

\begin{proposition}\label{PCMFailLocInv}
    The following captures the relationship between $\nu(\succ_x)$ and $\nu_{X \setminus \{x\}}(\succ_x)$ in the persistent craving model.
    \begin{equation}\label{PCMFailEq}
        \nu_{X \setminus \{x\}}(\succ_x)=\frac{\nu(\succ_x)}{1- \phi(M(\rhd,X \setminus \{x\}),x)(1-\nu(\succ_x))}
    \end{equation}
\end{proposition}

Proposition \ref{PCMFailLocInv} follows from Proposition \ref{GenFailLocInv} and tells us that there is a simple connection between $\nu(\succ_x)$, $\nu_{X \setminus \{x\}}(\succ_x)$, and $\phi(M(\rhd,X \setminus \{x\}),x)$. Notably, Equation \ref{PCMFailEq} reinforces the fact that the persistent craving model fails menu invariance only due to the persistence of cravings.

\subsection{A No Investment Condition}\label{A No Investment Condition}

In Section \ref{Local Invariance}, we studied local invariance and its relation to menu invariance. While local invariance has many appealing properties, one of its weaknesses is that it, in theory, requires us to know at least one stationary distribution $\nu_A$ before applying it as a test. Our goal in this section is to develop a characterization of menu invariance which does not require knowledge of any stationary distribution. This characterization will be through a no investment condition. The no investment condition can be thought of as an extension of the no trade condition of \citet{milgrom1982information} and \citet{morris1994trade} with the difference being that, instead of asking that there is no trade between agents, we ask that there is no investment between states (i.e. across time).

Our investment story proceeds as follows. Consider an agent who wants to develop an investment plan for each state of the world. This agents can invest at different banks and each of these banks has potentially differing forecasts for the future. Further, the agent is subject to a balanced budget or steady state constraint which means that the agent can only invest as much money at a bank as they already have at the bank. Lastly, this agent must be willing to maintain their investment plan in each state of the world, as, if they are not, the agent will be unable to commit to the investment plan. In our setup, a realized state of the world corresponds to a realized preference $\succ$, each bank corresponds to a choice set $A$, and the forecasts of each bank correspond to the transition probabilities of the Markov chain $M_A$. In order to understand this no investment condition, we first define an investment plan.

\begin{definition}
    A function $i:\mathcal{X} \times \mathcal{L}(X) \rightarrow \mathbb{R}^+$ is an \textbf{investment plan}. We call an investment plan a \textbf{strict investment plan} if it is not everywhere zero.
\end{definition}

An investment plan specifies how much our agent receives in each state of the world $\succ$ from each bank $A$. Suppose that $\succ$ is realized today. Given the balanced budget condition, the cost of investing today at bank $A$ is equal to $i(\succ,A)$ as our agent reinvests their entire return at each bank. We can also think about the expected revenue of investing at bank $A$ when today's state is $\succ$. We already know that bank $A$ pays out $i(\succ',A)$ when state $\succ'$ is realized. According to bank $A$, the likelihood that $\succ'$ is realized tomorrow, given that today's state is $\succ$, is given by $t_{\succ'}(M(\succ,A),\succ)$. Finally, given that our agent receives the payout from their investment tomorrow, the agent discounts the expected return of their investment at a rate of $\delta$. To summarize, the expected return on investing at bank $A$ in state $\succ$ is given by $\sum_{\succ' \in \mathcal{L}(X)}\delta i(A,\succ')t_{\succ'}(M(\succ,A),\succ)$. With all this in mind, we now introduce our no investment condition.

\begin{definition}\label{NoInvDef}
    We say that a transition function $t$ satisfies \textbf{no investment} if, for every strict investment plan and for every discount rate $\delta \in (0,1)$, there exists some $\succ \in \mathcal{L}(X)$ such that the following holds.
    \begin{equation}\label{NoInvEq}
        \underbrace{\sum_{(A,\succ') \in \mathcal{X} \times \mathcal{L}(X)} \delta i(A,\succ')t_{\succ'}(M(\succ,A),\succ)}_{\substack{\text{Expected revenue from investing} \\ \text{when $\succ$ is realized today}}} < \underbrace{\sum_{A \in \mathcal{X}}i(\succ,A)}_{\substack{\text{Total cost of investing} \\ \text{when $\succ$ is realized today}}}
    \end{equation}
\end{definition}

Simply put, no investment asks that, for every investment plan, there is always some state of the world $\succ$ where our agent would prefer to deviate from their investment plan.

\begin{theorem}\label{NoInvThm}
    A full support transition function $t$ is menu invariant if and only if it satisfies no investment.
\end{theorem}

Theorem \ref{NoInvThm} tells us that, subject to a balanced budget constraint, investment which is profitable in every state of the world is possible if and only if we have two (or more) banks whose forecasted stationary distribution differ. In terms of menu invariance, Theorem \ref{NoInvThm} acts as a test of menu invariance which requires no knowledge of any stationary distribution. An additional benefit of Theorem \ref{NoInvThm} is that it extends to an environment where the agent's choice set is allowed to vary over time. We consider such a setting in the appendix.

\section{Discussion}\label{Discussion}

In this paper we show that the presence of consumption dependence leads to the random utility model being a misspecified model of repeated rational choice. We characterize exactly when this misspecification occurs through local invariance and no investment. The natural followup question is how to deal with consumption dependence when we have either repeated cross-sectional data or standard market level data. These are both fruitful directions for future research and we believe that Proposition \ref{GenFailLocInv} is the first step in developing a tool kit for these types of data.

Before concluding with a discussion of the related literature, we first return to our motivating example in empirical industrial organization. Empiricists frequently use time period fixed effects in their static random utility models in order to capture dynamic heterogeneity. While this setup is not the focus of our paper, we show through a brief example that many of the problems discussed in this paper are still present even when time period fixed effects are used.

\begin{example}[Logit Habit Formation and Time Fixed Effects]\label{EX:fixedeffects}
    Consider a model of dynamic choice where each period's choice frequencies are governed by \begin{equation}\label{EQ:logitexample}
    p(x,A|y)=\frac{e^{v(x)+c(x) \mathbf{1}\{x=y\}}}{\sum_{z \in A} e^{v(z)+c(z) \mathbf{1}\{z=y\}}},
\end{equation} the conditional choice frequencies of the habit formation logit model. This corresponds to the true model. Suppose that an analyst has access to choice data from choice set $\{x,o\}$ at two discrete time periods. We assume that each agent in a population chooses once in each of these time periods. Suppose that the analyst is trying to estimate the model given by \begin{equation}\label{EQ:logitfixedEffects}
    p_t(x,A)= \frac{e^{v_x + f{(x,t)}}}{\sum_{y \in A}e^{v_y + f{(y,t)}}}
\end{equation} where $f(x,t)$ corresponds to a time period $t$ fixed effect on alternative $x$. At one extreme, we could consider the case of what happens when each agent's choices are in a steady state. In this case, the data from each period corresponds Equation \ref{HabitLogitStat}. This leads to the same biased estimator $\hat{v}$ in Equation \ref{biasedEstimator} with each fixed effect term being estimated equal to zero.

Now suppose we are in the other extreme case where in the first period, every agent chooses as if they chose $o$ in the previous period. For simplicity, we consider the following parameters: $v(o)=c(o)=v(x)=0$ and $c(x)=\log(2)$. This corresponds to data from when $x$ is first introduced to a market. In this case, our data corresponds to choosing $x$ half the time in the first period and with frequency $\frac{7}{12}$ in the second period. We consider three ways of estimating $v(x)$ in order to recover $f(x,t)$. In each case, we assume $f(o,1)=f(o,2)=0$.\begin{enumerate}
    \item Estimate $v(x)$ from the aggregate choice data (i.e. the average of $p_1(x,\{x,o\})$ and $p_2(x,\{x,o\})$) and then estimate $f(x,1)$ and $f(x,2)$
    \item Estimate $v(x)$ from $p_1(x,A)$ and then estimate $f(x,1)$ and $f(x,2)$
    \item Estimate $v(x)$ from $p_2(x,A)$ and then estimate $f(x,1)$ and $f(x,2)$
\end{enumerate}
We summarize the estimated values in each of these cases in Table \ref{tab:logitfixedeffects}. In the case where we estimate $\hat{v}(x)$ from the aggregate data, we see that our estimator is biased to be above the true value of $v(x)$. In the other two cases, we get estimators which are simply translations of each other. When period 1 data is used to estimate $\hat{v}(x)$, we actually get an unbiased estimator of $v$. However, as the model estimated by the analyst makes no effort to actually model the behavior leading to dynamic heterogeneity, the connection between $\hat{f}$ and $c$ is lost. Finally, this example appears to show that, when the only dynamic behavior of concern is habit formation, then estimating $\hat{v}(x)$ from data on the first time period $x$ is available leads to an unbiased estimator. This is unlikely to be the case once we entertain dynamic behaviors such as the incentive to experiment/explore new options which is likely to be present the first period a new product is made available. Further, we have made the assumption that each agent chooses a single time in each time period which is unlikely to be the case in markets, such as the beer or cereal market, when data is aggregated to the quarterly level. In summary, static models of random utility with time fixed effects seem to be subject to many of the same problems as those discussed in Section \ref{Examples}.

\begin{table}
    \centering
    \begin{tabular}{c|c|c|c}
        Estimating $\hat{v}$ from & $\hat{v}(x)$ & $\hat{f}(x,1)$ & $\hat{f}(x,2)$\\ \hline
        Aggregate data & $\log\left(\frac{13}{11}\right)$ & $\log\left(\frac{11}{13}\right)$  & $\log\left(\frac{7}{5}\right)-\log\left(\frac{13}{11}\right)$ \\
        Period 1 data & $0$ & $0$ & $\log\left(\frac{7}{5}\right)$ \\
        Period 2 data & $\log\left(\frac{7}{5}\right) $& $-\log\left(\frac{7}{5}\right)$ & $0$ \\
    \end{tabular}
    \caption{The estimator values for each case of the three estimation cases proposed in Example \ref{EX:fixedeffects}.}
    \label{tab:logitfixedeffects}
\end{table}
\end{example}

\subsection{Related Literature}
We now conclude with a discussion of the related literature. Our paper is most closely related to the work of \citet{lu2020repeated} which also studies the problem of aggregating repeated choice. Unlike us, they work with agents who are forward looking and subject to state dependence but are not consumption dependent. Their goal is to see if ignoring forward looking behavior causes bias in empirical exercises. They show that bias arises if and only if the agent's utility function does not take the form $U(c,v)=(1-\beta)u(c)+\beta v$ where $c$ is today's consumption and $v$ is the agent's continuation value. We view our results as complementary to those of \citet{lu2020repeated}. Their results cover forward looking behavior and our results cover backward looking behavior.

To our knowledge, there is very little work studying the empirical content of aggregated choice when agents are consumption dependent. \citet{valkanova2021markov} considers a model where agents have consumption dependent attention. In any given period, an agent will compare the alternative they just chose with a single other alternative according to some probability distribution. They make the assumption that the ratio of the transition probabilities from $x$ to $y$ and from $y$ to $x$ is constant across menus. This notably differs from the assumption that consumption dependent logit puts on transition probabilities. In consumption dependent logit, the transition probabilities from $x$ to $y$ and from $x$ to $z$ must have a constant ratio across menus. In general, consumption dependent random utility puts no restrictions on how the transition probabilities from $x$ and from $y$ compare but can put restrictions on how the transition probabilities from $x$ to $y$ and from $x$ to $z$ compare.\footnote{In the case of a Markovian version of state independent consumption dependent random utility, the transition probabilities from $x$ to $y$ must satisfy complete monotonicity. This in fact characterizes transition probabilities between alternatives consistent with Markovian state independent consumption dependent random utility.}

Our paper also contributes to the literature which studies dynamic extensions of the random utility model. Foundational to this literature are \citet{block1959random} and \citet{gul2006random}. The first axiomatically studies random utility and the second axiomatically studies random expected utility. To our knowledge, \citet{fudenberg2015dynamic} is the first to axiomatically study a dynamic random utility model. They study a type of dynamic logit model which is commonly used in dynamic discrete choice settings. \citet{frick2019dynamic} consider an extension of \citet{gul2006random} and \citet{ahn2013preference} and axiomatize a general nonparametric model of dynamic random expected utility. In a working version of their paper, they consider an extension of their base model which allows for consumption dependence. More recently, there has been work by \citet{li2021axiomatization}, \citet{chambers2021correlated}, and \citet{kashaev2022nonparametric} which studies agents who are subject to state dependence in a general abstract setting. \citet{turansick2024consumption} considers an extension of these models allowing for both consumption and state dependence and offers and axiomatization of this model when analysts have access to data on the frequency of consumption streams.

Beyond the work we have already mentioned on dynamic random utility, \citet{lu2018random} study intertemporal choice when the agent's discount rate is random. \citet{pennesi2021intertemporal} studies the difference between intertemporal Luce and logit models. The key difference between the two models is that the discount factor enters exponentially in the logit model while it does not in the Luce model. \citet{strack2021dynamic} consider a two period model. In the second period, the agent chooses according to a random utility. In the first period, the agent chooses according to an expected utility function. \citet{strack2021dynamic} study when the expected utility function in the first period can be induced by the random utility in the second period. \citet{deb2021dynamic} study a model of common learning. In this model, they study when the dynamic choices of a population of expected utility maximizers can be induced by a common stream of information.

\appendix

\section{Proofs for Section \ref{Examples}}

\subsection{Proofs from Section \ref{PersistenCravingsRCR}}

We actually begin with the proof of Proposition \ref{PCMMI} and use this result to prove Proposition \ref{PCMReg} and Proposition \ref{PCMComparative}.

\begin{proof}
 We actually prove something stronger than Proposition \ref{PCMMI}. We prove that $\nu_{X \setminus \{x\}}(\succ_x)>\nu_{X}(\succ_x)$. Once we observe that $\phi(x,y)\in(0,1)$ when $x\neq y$ and $\nu(\succ_x) < 1$, this follows immediately from Proposition \ref{PCMFailLocInv}, and so we are done.
\end{proof}

\begin{proof}
    We now proceed with our proof of Proposition \ref{PCMReg}. In the previous proof we showed that $\nu_{X \setminus \{x\}}(\succ_x)>\nu_{X}(\succ_x)$. Further, we make the observation that the support of $\nu$ will be the support of $\nu_A$ for all $A \in \mathcal{X}$. Now observe that for all $w,y,z \in X$ such that $y\neq x$ and $z \neq x$, we have $t_{\succ_y}(M(\succ_w,X\setminus \{x\}),\succ_w)/t_{\succ_z}(M(\succ_w,X\setminus \{x\}),\succ_w)=t_{\succ_y}(w,\succ_w)/t_{\succ_z}(w,\succ_w)$. Thus we have $\nu_X(\succ_y)/\nu_X(\succ_z)=\nu_{X\setminus\{x\}}(\succ_y)/\nu_{X\setminus\{x\}}(\succ_z)$. We now know that $\nu_{X \setminus \{x\}}(\succ_x)>\nu_{X}(\succ_x)$ and that the ratio between $\nu_A(\succ_y)$ and $\nu_A(\succ_z)$ is constant between $X$ and $X\setminus\{x\}$ for all other $y$ and $z$. It immediately follows that $\nu_{X \setminus \{x\}}(\succ_y) < \nu_X(\succ_y)$ for all $y \in X \setminus \{x\}$. Let $z \neq M(\rhd,X \setminus \{x\})$ be some alternative in $X \setminus \{x\}$. $z$ exists as $|X|\geq 3$. It then follows that $p(z,X\setminus\{x\})=\nu_{X \setminus \{x\}}(\succ_z)<\nu_X(\succ_z)=p(z,X)$. Thus we have a failure of regularity, and so we are done.
\end{proof}

\begin{proof}
    We now prove Proposition \ref{PCMComparative}. By Proposition \ref{PCMFailLocInv}, we know that $\nu_{X\setminus \{x\}}(\succ_x)> \nu'_{X\setminus \{x\}}(\succ_x)$ if and only if $\phi(M(\rhd,X \setminus \{x\}),x)>\phi'(M(\rhd,X \setminus \{x\}),x)$. It then follows from the observation that $\frac{\nu_{X\setminus \{x\}}(\succ_y)}{\nu_{X\setminus \{x\}}(\succ_z)}=\frac{\nu_{X}(\succ_y)}{\nu_{X}(\succ_z)}$ for $y,z \in X \setminus \{x\}$ for $y \neq M(\rhd,X \setminus \{x\})$ and $z \neq M(\rhd,X \setminus \{x\})$ that a larger $\nu_{X\setminus \{x\}}(\succ_x)$ equates to a smaller $\nu_{X\setminus \{x\}}(\succ_y)$, and so we are done.
\end{proof}

\subsection{Proofs from Section \ref{HabitFormationRCR}}

We begin with a preliminary lemma.

\begin{lemma}\label{LogitHabFormStationary}
    In the habit formation logit model, the stationary distribution over alternatives in menu $A$ is dictated by the following equation.
    \begin{equation}
        p(x,A)=\frac{e^{v(x)}\left(\sum_{y\in A} e^{v(y) + c(y)\mathbf{1}\{x=y\}}\right)}{\sum_{z \in A}e^{v(z)}\left(\sum_{y\in A} e^{v(y) + c(y)\mathbf{1}\{z=y\}}\right)}
    \end{equation}
\end{lemma}

\begin{proof}
    We now verify our above statement.
    \begin{equation}
        \begin{split}
            p(x,A) & = \sum_{y \in A}p(y,A)p(x,A|y) \\
            & =\sum_{y \in A}\left(\frac{e^{v(y)}\left(\sum_{z\in A} e^{v(z) + c(z)\mathbf{1}\{y=z\}}\right)}{\sum_{w \in A}e^{v(w)}\left(\sum_{z\in A} e^{v(z) + c(z)\mathbf{1}\{w=z\}}\right)}\right)\left(\frac{e^{v(x)+c(x) \mathbf{1}\{x=y\}}}{\sum_{z \in A} e^{v(z)+c(z) \mathbf{1}\{z=y\}}}\right) \\
            & = \sum_{y \in A}\frac{e^{v(y)}e^{v(x)+c(x) \mathbf{1}\{x=y\}}}{\sum_{w \in A}e^{v(w)}\left(\sum_{z\in A} e^{v(z) + c(z)\mathbf{1}\{w=z\}}\right)} \\
            & =\sum_{y \in A}\frac{e^{v(x)}e^{v(y)+c(y) \mathbf{1}\{x=y\}}}{\sum_{w \in A}e^{v(w)}\left(\sum_{z\in A} e^{v(z) + c(z)\mathbf{1}\{w=z\}}\right)} \\
            & =\frac{e^{v(x)}\left(\sum_{y\in A} e^{v(y) + c(y)\mathbf{1}\{x=y\}}\right)}{\sum_{w \in A}e^{v(w)}\left(\sum_{z\in A} e^{v(z) + c(z)\mathbf{1}\{w=z\}}\right)}
        \end{split}
    \end{equation}
    Above, the first equality holds due to the definition of a stationary distribution. The second equality holds via substitution. The third equality holds due to canceling like terms. The fourth equality holds due to the fact that $e^{r+s}=e^r e^s$ and the fact that $c(x)=c(y)$ when $x=y$. The fifth equality holds by moving the sum two the numerator. This final term is exactly what we posited the stationary distribution to be, and so we are done. 
\end{proof}

\subsubsection{Proof of Proposition \ref{failIIA}}
\begin{proof}
    By Lemma \ref{LogitHabFormStationary}, we know that the following two expressions holds.
    \begin{equation}\label{iia1}
        \frac{p(x,\{x,o\})}{p(o,\{x,o\})}=\frac{e^{v(x)}(1+e^{v(x)+c(x)})}{1+e^{v(x)}}
    \end{equation}
    \begin{equation}\label{iia2}
        \frac{p(x,\{x,y,o\})}{p(o,\{x,y,o\})}=\frac{e^{v(x)}(1+e^{v(x)+c(x)}+e^{v(y)})}{1+e^{v(x)}+e^{v(y)}}
    \end{equation}
    The right hand sides of Equations \ref{iia1} and \ref{iia2} are equal if and only if $c(x)=0$, and so we are done.
\end{proof}

\subsubsection{Proof of Proposition \ref{biasedEstimator}}
\begin{proof}
    We begin by restating the standard estimator.
    \begin{equation*}
        \hat{v}(x)=\log\left(\frac{p(x,\{x,o\})}{p(o,\{x,o\})}\right)
    \end{equation*}
    We now use Lemma \ref{LogitHabFormStationary} to substitute.
    \begin{equation}
        \begin{split}
            \log\left(\frac{p(x,\{x,o\})}{p(o,\{x,o\})}\right) & = \log\left(\frac{e^{v(x)}(1+e^{v(x)+c(x)})}{1+e^{v(x)}}\right) \\
            & = v(x) + \log(1+e^{v(x)+c(x)}) - \log(1+e^{v(x)})
        \end{split}
    \end{equation}
    The last line above is equal to $v(x)$ if and only if $c(x)=0$, and so we are done.
\end{proof}

\section{Proofs for Section \ref{Menu Invariance}}

In this appendix, we extend some of our results from Section \ref{Menu Invariance} to transition functions which are not full support. To do so, we first need an updated definition of menu invariance. This new definition relies on the idea of an invariant distribution rather than a stationary distribution. An invariant distribution $\nu$ of a Markov chain with transition matrix $M$ is any distribution satisfying $\nu M=\nu$. When $M$ is ergodic, there is a unique invariant distribution which coincides with the Markov chain's stationary distribution. When $M$ is not ergodic, it will generally have (uncountably) many invariant distributions.

\begin{definition}\label{MIDefFull}
    We say that a transition function $t$ is \textbf{menu invariant} if there exists some $\nu \in \Delta(\mathcal{L}(X))$ such that $\nu M_A=\nu$ for all $A \in \mathcal{X}_2$.
\end{definition}

We use the above definition of menu invariance when discussing transition functions which are not full support.

\subsection{Proof of Proposition \ref{SIMenuInv}}
\begin{proof}
    If $t(x)=t(y)$ for all $x,y\in X$, then $t(x)$ defines the stationary distribution over preferences which is common to each choice set. Now suppose to a contradiction $t(x) \neq t(y)$ for some $x,y \in X$ and that $t$ is menu invariant with common stationary distribution $\nu$. Consider the set $\{x,y\}=A$. With a slight abuse of notation, let $t(x,y) = \sum_{\succ \in N(x,\{x,y\})}t_{\succ}(x)$. Define $t(y,x)$ similarly. It then follows that $t$ defines a Markov chain between choices on $A$ which is given by the following matrix.
\begin{equation*}
    \begin{bmatrix}
        1 - t(x,y) & t(x,y) \\
        t(y,x) & 1 - t(y,x)
    \end{bmatrix}
\end{equation*}

The first row and column of this matrix corresponds to $x$ and the second row and column corresponds to $y$. The stationary distribution of this Markov chain is $\begin{bmatrix}
    \frac{t(y,x)}{t(y,x) + t(x,y)} \\
    \frac{t(x,y)}{t(y,x) + t(x,y)}
\end{bmatrix}$. Since $t$ is strictly positive in each of its elements we can write $p(x,A)=\alpha$ and $p(y,A)=1-\alpha$ for some $\alpha \in (0,1)$. It then follows that $\nu = \alpha t(x) + (1-\alpha) t(y)$. Since $t(x) \neq t(y)$, there is some $\succ$ such that, without loss of generality, $t_\succ(x) < \nu(\succ) < t_\succ(y)$. Consider $z \not \in \{x,y\}$. Since the stationary distribution for each choice set is equal to $\nu$, it must be the case that $t_\succ(x) < \nu(\succ) < t_\succ(z)$ and $t_\succ(z) < \nu(\succ) < t_\succ(y)$ which is a contradiction, and so we are done.
\end{proof}

\subsection{Proof of Theorem \ref{LocalInvThm}}
In no part of the following proof do we rely on having a full support transition function. As such, Theorem \ref{LocalInvThm} holds for general transition functions using the definition of menu invariance introduced earlier in this appendix.
\begin{proof}
    Our proof strategy will be to show that $(1) \implies (2) \implies (3) \implies (1)$. It is obvious that that $(2)$ implies $(3)$, so all we have left to do is to prove $(1)$ implies $(2)$ and $(3)$ implies $(1)$. We begin by showing $(1)\implies (2)$. Suppose that $t$ is menu invariant and let $x \in A$ be such that $|A|\geq3$. We then have the following for some $\nu_B$ in the set of invariant distributions of $M_B$.
    \begin{equation}\label{LocInvPf}
        \begin{split}
            & \sum_{\succ \in \mathcal{L}(X)}\nu_B(\succ)t(M(\succ,A),\succ)-\sum_{\succ\in\mathcal{L}(X)}\nu_B(\succ)t(M(\succ,A\setminus\{x\}),\succ) \\
            &=\sum_{\succ \in N(x,A)} \nu_B(\succ)t(x,\succ) - \sum_{y \in A \setminus \{x\}} \sum_{\succ \in N(x,A) \cap N(y,A \setminus \{x\})}  \nu_B(\succ)t(y,\succ)
        \end{split}
    \end{equation}
    Above, the left hand side of the equality represents $\nu_B[M_A - M_{A\setminus\{x\}}]$. The equality between the two sides holds by gathering like terms. The left hand side is equal to zero as $t$ is menu invariant. This then means that the right hand side of the equality is equal to zero for arbitrary $A$. Thus $t$ is locally invariant with respect to $\nu_B$. Since $\nu_B$ was also chosen arbitrarily, $t$ is locally invariant with respect to $\nu_B$ for all $B \in \mathcal{X}$

    We now show $(3)\implies (1)$. Suppose that $t$ is locally invariant with respect to $\nu_A$. As we showed in Equation \ref{LocInvPf}, subtracting the two terms in Equation \ref{LocalInvDef} is equal to $\nu[M_A - M_{A\setminus\{x\}}]=\nu[(M_A-I) - (M_{A\setminus\{x\}}-I)]$ where $I$ is the identity matrix. As $\nu_A$ is the stationary distribution of $M_A$, it follows that $\nu_A(M_A-I)=0$. As $t$ is locally invariant with respect to $\nu_A$ it then follows that $\nu_A[M_A - M_{A\setminus\{x\}}]=0$ and $\nu_A(M_{A\setminus\{x\}}-I)=0$. Thus $\nu_A$ is the stationary distribution for $M_{A\setminus\{x\}}$. We can repeat this argument for $\nu_A[M_{A\cup\{x\}}-M_A]$ and get that $\nu_A$ is the stationary distribution for $M_{A\cup\{x\}}$. It then follows from iterative application of the previous argument that $\nu_A$ is the stationary distribution of $M_X$. These arguments can once again be applied iteratively to argue that for any $B \subseteq X$ with $|B|\geq 2$ that $\nu_A$ is the stationary distribution of $M_B$. Thus $t$ is menu invariant, and so we are done.
\end{proof}
\subsection{Proof of Propositions from Section \ref{Failures of LI}}

\subsubsection{Proof of Proposition \ref{2eLocFail}}
\begin{proof}
    
It immediately follows from the definition of $\epsilon_A$ and $\epsilon_{A\setminus \{x\}}$ that $\epsilon_A-\epsilon_{A\setminus\{x\}}=(\nu_A-\nu_{A\setminus\{x\}})[M_A-M_{A \setminus \{x\}}]$ is true. We then proceed to the proof of the second statement. A property of the Moore-Penrose inverse of a matrix $M$ is that when $M$ has full rank $M M^{mp}$ is equal to the identity matrix. We then get that $(\epsilon_A-\epsilon_{A\setminus\{x\}})[M_A-M_{A \setminus \{x\}}]^{mp}=\nu_A-\nu_{A\setminus\{x\}}$ by right multiplying by the Moore-Penrose inverse of $[M_A-M_{A \setminus \{x\}}]$ in our prior equation.
\end{proof}
\subsubsection{Proof of Proposition \ref{GenFailLocInv}}
\begin{proof}

Equation \ref{GenFailLocInvEq} follows by substituting our notation into Theorem 2.3 of \citet{hunter2005stationary}. To show Equation \ref{FailLocInvSingleEq}, we begin by multiplying the right hand side of \ref{GenFailLocInvEq} and focusing on a single entry of the resulting vector.
\begin{equation}
    \begin{split}
        \nu_{A}(\succ)-\nu_{A\setminus \{x\}}(\succ)& =\sum_{\succ'\neq \succ}\epsilon_A(\succ')(-n_{A\setminus \{x\}}(\succ',\succ))(1/n_{A\setminus \{x\}}(\succ,\succ)) \\
        & =-\sum_{\succ'\neq \succ}\epsilon_A(\succ')n_{A\setminus \{x\}}(\succ',\succ)\nu_{A\setminus \{x\}}(\succ)
    \end{split}
\end{equation}
The first equality holds looking at a single entry of Equation \ref{GenFailLocInvEq}. The second equality holds as the weight the stationary distribution puts on a state in an ergodic Markov chain is equal to the inverse of the mean return time of that state. We then get the following.
\begin{equation}\label{Proposition7almostdone}
    \nu_A(\succ) = \nu_{A\setminus \{x\}}(\succ)[1-\sum_{\succ' \neq \succ}\epsilon_{A}(\succ')n_{A\setminus \{x\}}(\succ',\succ)]
\end{equation}
Once we divide both sides of Equation \ref{Proposition7almostdone} by the bracketed value, we are left with Equation \ref{FailLocInvSingleEq}, and so we are done.

\end{proof}

\subsubsection{Proof of Proposition \ref{PCMFailLocInv}}
\begin{proof}
    Observe the following.
    \begin{equation*}
        \begin{split}
            \nu_{X \setminus \{x\}}(\succ_x) & = \frac{\nu_X(\succ_x)}{1 - \sum_{y \neq x}\epsilon_{X}(\succ_y)n_{A \setminus \{x\}}(\succ_y,\succ_x)} \\
            & = \frac{\nu(\succ_x)}{1 - \sum_{y \neq x}\nu(\succ_x)[\nu(\succ_y)-(1-\phi(M(\rhd,X \setminus \{x\}),x))\nu(\succ_y)]/\nu(\succ_x)} \\
            & = \frac{\nu(\succ_x)}{1 - \sum_{y \neq x}\nu(\succ_y)\phi(M(\rhd,X \setminus \{x\}),x))} \\
            & = \frac{\nu(\succ_x)}{1 - (1-\nu(\succ_x))\phi(M(\rhd,X \setminus \{x\}),x))} \\
        \end{split}
    \end{equation*}
    Above, the first line follows directly from Proposition \ref{GenFailLocInv}. The second line follows from direct calculation of $\epsilon_X(\succ_y)$, the fact that the mean passage time from $y \neq x$ to $x$ is the same at $X$ and $X \setminus \{x\}$, the fact that $t(y,\succ_y)$ is i.i.d., and that $\nu(\succ_x)=1/n_X(\succ_X,\succ_x)$. The third line follows from collecting like terms. The last line holds as $\sum_{z \in X}\nu(\succ_z)=1$, and so we are done.
\end{proof}
\subsection{Proof of Theorem \ref{NoInvThm}}

We begin by offering an extended definition of no investment.

\begin{definition}\label{NoInvDefFull}
    We say that a transition function $t$ satisfies \textbf{no weak investment} if, for every investment plan, there exists some $\succ \in \mathcal{L}(X)$ such that the following holds.
    \begin{equation}\label{NoInvEqFull}
        \underbrace{\sum_{(A,\succ') \in \mathcal{X} \times \mathcal{L}(X)} i(A,\succ')t_{\succ'}(M(\succ,A),\succ)}_{\substack{\text{Expected revenue from investing} \\ \text{when $\succ$ is realized today}}} \leq \underbrace{\sum_{A \in \mathcal{X}}i(\succ,A)}_{\substack{\text{Total cost of investing} \\ \text{when $\succ$ is realized today}}}
    \end{equation}
\end{definition}

No weak investment differs from no investment as no weak investment allows for investment plans which are not strict but weakens the strict equality from no investment to a weak inequality. This no weak investment condition allows us to deal with general transition functions.

\begin{theorem}\label{NoInvThmFull}
    A transition function $t$ is menu invariant if and only if it satisfies no weak investment.
\end{theorem}

We begin by proving Theorem \ref{NoInvThmFull}. The proof of Theorem \ref{NoInvThm} then follows as a corollary.

\begin{proof}
    The condition for a distribution to be stationary is $\nu(M - I) = 0$ where $I$ is the identity matrix. The condition for menu invariance can be written similarly. Consider the matrix $M$ with rows indexed by elements of $\mathcal{L}(X)$ and columns indexed by elements of $\mathcal{X}_2 \times \mathcal{L}(X)$. The typical element of $M$ is given as follows.
\begin{equation*}
    m(\succ,(A,\succ')) = t_{\succ'}(M(\succ,A),\succ)-\mathbf{1}\{\succ = \succ'\}
\end{equation*}
Menu invariance can now be written as the existence of a $\nu$ such that $\nu M = 0$. Ville's Theorem of the Alternative (see \citet{ville1938theorie} and \citet{borderAlternative}) tells us that there exists $\nu > 0$ such that $\nu M \leq 0$ if and only if there does not exist some $i \geq 0$ such that $Mi >> 0$. Consider some $\nu>0$ such that $\nu M\leq 0$. If such a $\nu$ exists, then we can rescale $\nu$ to be a probability distribution (i.e. $\nu \cdot 1=1$). $M$ is written as a series of $(M_A- I)$ where $M_A$ is the Markov transition matrix for choice set $A$ in Section \ref{Model}. As such $\nu M_A \cdot 1 = \nu I \cdot 1 = 1$. This means that $\nu (M_A - I) \cdot 1 =0$ which in turn tells us that $\nu M  \cdot 1 = 0$. Further, if there is some component of $\nu M$ that is strictly less than zero then there must be some other component of $\nu M$ strictly larger than zero. Finally, this gives us that $\nu M \leq 0$ and $\nu > 0$ if and only if $\nu M = 0$ and $\nu > 0$.  

Now suppose there exists some $i \geq 0$ such that $Mb >>0$. This $i$ is an investment plan from our no investment condition. If we write out each inequality implied by $Mi >>0$ we get the following for each $\succ \in \mathcal{L}(X)$.
\begin{equation*}
    \sum_{(A,\succ') \in \mathcal{X}_2 \times \mathcal{L}(X)}i(A,\succ')[t_{\succ'}(M(\succ,A),\succ) - \mathbf{1}\{\succ = \succ'\} ]> 0
\end{equation*}

Note that $\sum_{(A,\succ') \in \mathcal{X}_2 \times \mathcal{L}(X)}i(A,\succ')\mathbf{1}\{\succ = \succ'\}$ is exactly $\sum_{A \in \mathcal{X}_2}i(\succ,A)$. This means we can rewrite the above as follows.
\begin{equation*}
    \sum_{(A,\succ') \in \mathcal{X}_2 \times \mathcal{L}(X)}i(A,\succ')t_{\succ'}(M(\succ,A),\succ) - \sum_{A\in \mathcal{X}_2}i(\succ,A) >0
\end{equation*}
The negation of the above holding for all $\succ$ is exactly our no investment condition. So our no investment condition holds if and only if there does not exist some $i \geq 0$ with $Mi >> 0$. This is equivalent to the existence of some $\nu > 0$ satisfying $\nu \cdot 1 = 1$ and $\nu M \leq 0$ by Ville's Theorem of the Alternative. Finally we showed that this is equivalent to the existence of some $\nu > 0$ satisfying $\nu \cdot 1 = 1$ and $\nu M = 0$ which is exactly menu invariance, and so we are done.
\end{proof}

We now proceed with our proof of Theorem \ref{NoInvThm}.

\begin{proof}
    Note that in the proof of Theorem \ref{NoInvThmFull}, $\mathbf{0}$ can never be the vector which causes no weak investment to fail. It then follows that no weak investment is equivalent to no weak investment restricted to strict investment plans. We are now working with full support transition functions. This means that the left hand side of Equations \ref{NoInvEqFull} is always strictly positive when restricted to strict investment plans. So the condition from Theorem \ref{NoInvThmFull} holds if and only if we have some strictly positive number being less than or equal to some other strictly positive number for all 
    $\succ$. A strictly positive number $n$ is weakly less than strictly positive number $m$ if and only if $\delta n$ is strictly less than $m$ for all $\delta \in (0,1)$. This shows that Definition \ref{NoInvDef} and Definition \ref{NoInvDefFull} are equivalent when restricted to strict investment plans, and so we are done.
\end{proof}

\section{When Choice sets Vary}\label{Choice Sets Vary}

Thus far we have made the assumption that the agent's choice set is fixed over time. In most settings, this is an unreasonable assumption. In this section, we extend the model introduced in Section \ref{Model} by allowing the agent's choice set to vary exogenously over time according to a Markov chain. Our goal is to characterize the analogue of menu invariance in this setting via an extension of Theorem \ref{NoInvThm}. Before describing this extended model, we first define arrival functions which define the Markov chain over sets.

\begin{definition}
    We call a function $s:\mathcal{X} \rightarrow \Delta(\mathcal{X})$ an \textbf{arrival function}. Further, we call a function $s:\mathcal{X} \rightarrow int\Delta(\mathcal{X})$ a \textbf{full support arrival function}. 
\end{definition}

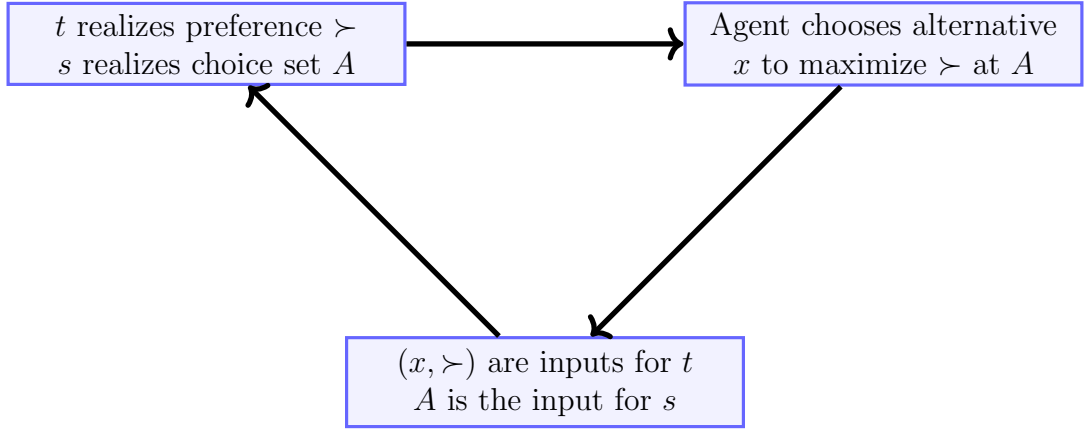
\begin{figure}
    \centering
    \begin{tikzpicture}[set/.style={rectangle, draw=blue!60, fill=blue!5, very thick, minimum size=7mm},
choice/.style={diamond, draw=red!60, fill=red!5, very thick, minimum size=7mm},
attention/.style={rectangle, draw=green!60, fill=green!5, very thick, minimum size=7mm}]

        \node[set,text width=5cm, align=center] (h) at (0,0) {$t$ realizes preference $\succ$ \\ $s$ realizes choice set $A$};
        \node[set,text width=5cm,align=center] (i) at (4.5,-4.5) {$(x,\succ)$ are inputs for $t$ \\ $A$ is the input for $s$};
        \node[set,text width=5cm,align=center] (x) at (9,0) {Agent chooses alternative \\ $x$ to maximize $\succ$ at $A$};

        \draw [->,line width=2pt] (i) -- (h) node[midway, below, sloped, scale=.8] {};
        \draw [->,line width=2pt] (h) -- (x) node[midway, below, sloped, scale=.8] {};
        \draw [->,line width=2pt] (x) -- (i) node[midway, below, sloped, scale=.8] {};

    \end{tikzpicture}
    \caption{A visual representation of the data generating process when choice sets are allowed to vary over time. At the start of a period, a preference $\succ$ and a choice set $A$ are realized. The agent then chooses $x$ to maximize $\succ$. Next period's preference is then determined by $t(x,\succ)$ and next period's choice set is determined by $s(A)$.}
    \label{fig:DGP2}
\end{figure}

We use the notation $s_B(A)$ to denote the probability that tomorrow's choice set is $B$ given that today's choice set is $A$. The data generating process of our extended model proceeds as follows. At the start of a period, a preference $\succ$ and a choice set $A$ are realized. The agent chooses $M(\succ,A)$ to maximize their preference. Then $t(M(\succ,A),\succ)$ determines next period's preference and $s(A)$ determines next period's choice set. Figure \ref{fig:DGP2} offers a visual representation of the extended model's data generating process.

We are interested in a setting where tomorrow's choice set is independent of today's choice. As such, we consider the independent mixture of our arrival function $s$ and our transition function $t$. Notably, a transition function $t$ and arrival function $s$ jointly define a Markov chain with states of the form $(\succ,A)$. The transition probability from $(\succ,A)$ to $(\succ',B)$ is given by $s_B(A)t_{\succ'}(M(\succ,A),\succ)$. We use $\pi$ to denote a typical element of $\Delta(\mathcal{X})$ and $\psi$ to denote a typical element of $\Delta(\mathcal{X}\times\mathcal{L}(X))$. Unlike in our base model, in the extended model each choice set does not have its own Markov chain over preferences. As such, we need to update our definition of menu invariance. 

\begin{definition}
    We say that a full support transition function $t$ and a full support arrival function $s$ are \textbf{jointly menu invariant} if the unique stationary distribution $\psi$ of the Markov chain defined by $t$ and $s$ can be written as $\psi(A,\succ)=\pi(A)\nu(\succ)$ for some $\pi \in \Delta(\mathcal{X})$ and some $\nu \in \Delta(\mathcal{L}(X))$.
\end{definition}

Joint menu invariance asks that the stationary distribution $\psi$ can be written as an independent mixture of the marginal stationary distribution over preferences and the marginal stationary distribution over choice sets. In Theorem \ref{NoInvThm}, our no investment condition made no reference to how choice sets vary over time as we had yet to add that to the model. In order to extend Theorem \ref{NoInvThm} to this extended model, we must encode some information about $s$ into our no investment condition. Given the stationary distribution over choice sets $\pi$ for arrival function $s$, we can define the probability that last period's choice set is $B$ given that this period's choice set is $A$. We will denote this probability using $\rho_s(B|A)$.
\begin{equation}\label{ArrivalYesterdayProb}
    \rho_s(B|A)=\frac{\pi(B)s_A(B)}{\sum_{C \in \mathcal{X}_2}\pi(C)s_A(C)}=\frac{\pi(B)s_A(B)}{\pi(A)}
\end{equation}
Equation \ref{ArrivalYesterdayProb} lets us extend our definition of no investment to this setting of menu variation.
\begin{definition}\label{NoInvDefSetVary}
    We say that a transition function $t$ and arrival function $s$ satisfy \textbf{no investment} if, for every strict investment plan and for every discount rate $\delta \in (0,1)$, there exists some $\succ \in \mathcal{L}(X)$ such that the following holds.
    \begin{equation}\label{NoInvEqSetVary}
        \underbrace{\sum_{(A,\succ') \in \mathcal{X} \times \mathcal{L}(X)} \sum_{B \in \mathcal{X}} \delta i(A,\succ')\rho_s(B|A)t_{\succ'}(M(\succ,B),\succ)}_{\substack{\text{Expected revenue from investing} \\ \text{when $\succ$ is realized today}}} < \underbrace{\sum_{A \in \mathcal{X}}i(\succ,A)}_{\substack{\text{Total cost of investing} \\ \text{when $\succ$ is realized today}}}
    \end{equation}
\end{definition}

Just as Equation \ref{NoInvEq} did, Equation \ref{NoInvEqSetVary} has a no investment interpretation. The notable difference between these two equations is that we have added $\rho_s(B|A)$ to the expected revenue side of Equation \ref{NoInvEqSetVary}. The addition of $\rho_s(B|A)$ follows from the fact that when our agent chooses from $A$ today, they face choice set $B$ with positive probability tomorrow. This further means that the stationary distribution over preferences at choice set $B$ depends on the stationary distribution over preferences at choice set $A$ as well as the transition probabilities of $M_A$. We now conclude this section with our extension of Theorem \ref{NoInvThm}.

\begin{theorem}\label{NoInvChoiceSetVaryThm}
    A full support transition function $t$ and full support arrival function $s$ are jointly menu invariant if and only if they satisfy no investment.
\end{theorem}

\begin{proof}
    Our proof of this result will proceed very much in a similar way to our proof of Theorem \ref{NoInvThm}. To begin, consider the following equation.
        \begin{equation}\label{NoInvEqWeak}
        \sum_{(A,\succ') \in \mathcal{X} \times \mathcal{L}(X)} \sum_{B \in \mathcal{X}}  i(A,\succ')\rho_s(B|A)t_{\succ'}(M(\succ,B),\succ)\leq \sum_{A \in \mathcal{X}}i(\succ,A)
    \end{equation}
    Equation \ref{NoInvEqWeak} is the analogue of Equation \ref{NoInvEqWeak} in the environment of Theorem \ref{NoInvChoiceSetVaryThm}. As argued in the proof of Theorem \ref{NoInvThm}, having $i=0$ can never cause the Equation \ref{NoInvEqWeak} to fail, so Equation \ref{NoInvEqWeak} holding for all investment plans is equivalent to Equation \ref{NoInvEqWeak} holding for all strict investment plans. As prior, when we have a strict investment plan, a full support transition function, and a full support arrival function, both sides of the inequality are strictly positive. It then follows from the arguments in the proof of Theorem \ref{NoInvThm} that Equation \ref{NoInvChoiceSetVaryThm} holds for all $\succ$ if and only if our no investment condition holds.

     Consider the matrix $M$ with rows indexed by elements of $\mathcal{L}(X)$ and columns indexed by elements of $\mathcal{X} \times \mathcal{L}(X)$. The typical element of $M$ is given as follows.
\begin{equation*}
    m(\succ,(A,\succ')) = \sum_{B \in \mathcal{X}}\rho_s(B|A)t_{\succ'}(M(\succ,B),\succ)-\mathbf{1}\{\succ = \succ'\}
\end{equation*}
As prior, Ville's Theorem of the Alternative tells us that there exists $\nu > 0$ such that $\nu M \leq 0$ if and only if there does not exist some $i \geq 0$ such that $Mi >> 0$. Using an analogous argument from the proof of Theorem \ref{NoInvThmFull}, we can show that $\nu M \leq 0$ and $\nu > 0$ if and only if $\nu M = 0$ and $\nu > 0$. Now suppose there exists some $i \geq 0$ such that $Mi >>0$. This $i$ is an investment plan from our no investment condition. If we write out each inequality implied by $Mi >>0$ we get the following for each $\succ \in \mathcal{L}(X)$.
\begin{equation*}
    \sum_{(A,\succ') \in \mathcal{X} \times \mathcal{L}(X)}i(A,\succ')[\sum_{B \in \mathcal{X}}\rho_s(B|A)t_{\succ'}(M(\succ,B),\succ)-\mathbf{1}\{\succ = \succ'\} ]> 0
\end{equation*}
As before, we can rewrite the above as follows.
\begin{equation*}
    \sum_{(A,\succ') \in \mathcal{X} \times \mathcal{L}(X)}\sum_{B \in \mathcal{X}}i(A,\succ')\rho_s(B|A)t_{\succ'}(M(\succ,B),\succ) - \sum_{A\in\mathcal{X}}i(\succ,A) >0
\end{equation*}
This is the negation of our no money pump condition. As prior this tells us that our no money pump condition holds if and only if there exists some $\nu > 0$ satisfying $\nu \cdot 1 = 1$ and $\nu M = 0$. Unlike prior, we are not done. 

Our goal now is to show that the $\nu$ we just found corresponds to $\pi \cdot \nu$ in the stationary distribution of our initial Markov chain formed by $s$ and $t$. We now verify that $\pi \nu$ is the stationary distribution whenever no money pump holds.
\begin{equation*}
    \begin{split}
        & \sum_{(\succ,B) \in \mathcal{L}(X)\times \mathcal{X}}\pi(B)\nu(\succ)t_{\succ'}(M(\succ,B),\succ)s_A(B) \\
        & = \pi(A)\sum_{(\succ,B) \in \mathcal{L}(X)\times \mathcal{X}}\rho_s(B|A)\nu(\succ)t_{\succ'}(M(\succ,B),\succ) \\
        & =\pi(A)\nu(\succ)
    \end{split}
\end{equation*}
The first line above is the probability of $(\succ,A)$ in the next period given that the distribution this period is given by $\pi\nu$. The second line follows from multiplying by $\frac{\pi(A)}{\pi(A)}$ and then collecting like terms to write $\rho_s(B|A)$. The last line then follows from the fact that $\nu M=0$. Notably the last line holds for all $(\succ,A)$ if and only if we have $\nu M =0$ which is true if and only if our no investment condition holds. The equality above shows that $\pi \cdot \nu$ is the stationary distribution if and only if no investment holds, and so we are done.

\end{proof}

\bibliographystyle{ecta}
\bibliography{mrum}

\begin{thebibliography}{38}
\newcommand{\enquote}[1]{``#1''}
\expandafter\ifx\csname natexlab\endcsname\relax\def\natexlab#1{#1}\fi

\bibitem[\protect\citeauthoryear{Ahn and Sarver}{Ahn and Sarver}{2013}]{ahn2013preference}
\textsc{Ahn, D.~S. and T.~Sarver} (2013): \enquote{Preference for Flexibility and Random Choice,} \emph{Econometrica}, 81, 341--361.

\bibitem[\protect\citeauthoryear{Block and Marschak}{Block and Marschak}{1959}]{block1959random}
\textsc{Block, H.~D. and J.~Marschak} (1959): \enquote{Random Orderings and Stochastic Theories of Response,} Tech. rep., Cowles Foundation for Research in Economics, Yale University.

\bibitem[\protect\citeauthoryear{Border}{Border}{2013}]{borderAlternative}
\textsc{Border, K.} (2013): \enquote{Alternative Linear Inequalities,} California Institute of Technology.

\bibitem[\protect\citeauthoryear{Carrasco, Labeaga, and David L{\'o}pez-Salido}{Carrasco et~al.}{2005}]{carrasco2005consumption}
\textsc{Carrasco, R., J.~M. Labeaga, and J.~David L{\'o}pez-Salido} (2005): \enquote{Consumption and habits: evidence from panel data,} \emph{The Economic Journal}, 115, 144--165.

\bibitem[\protect\citeauthoryear{Carroll, Overland, and Weil}{Carroll et~al.}{2000}]{carroll2000saving}
\textsc{Carroll, C.~D., J.~Overland, and D.~N. Weil} (2000): \enquote{Saving and growth with habit formation,} \emph{American Economic Review}, 90, 341--355.

\bibitem[\protect\citeauthoryear{Chambers, Masatlioglu, and Turansick}{Chambers et~al.}{2024}]{chambers2021correlated}
\textsc{Chambers, C.~P., Y.~Masatlioglu, and C.~Turansick} (2024): \enquote{Correlated choice,} \emph{Theoretical Economics}, 19, 1087--1117.

\bibitem[\protect\citeauthoryear{Chen and Risen}{Chen and Risen}{2010}]{chen2010choice}
\textsc{Chen, M.~K. and J.~L. Risen} (2010): \enquote{How choice affects and reflects preferences: revisiting the free-choice paradigm.} \emph{Journal of personality and social psychology}, 99, 573.

\bibitem[\protect\citeauthoryear{Deb and Renou}{Deb and Renou}{2021}]{deb2021dynamic}
\textsc{Deb, R. and L.~Renou} (2021): \enquote{Dynamic Choices and Common Learning,} \emph{arXiv preprint arXiv:2105.03683}.

\bibitem[\protect\citeauthoryear{Frick, Iijima, and Strzalecki}{Frick et~al.}{2019}]{frick2019dynamic}
\textsc{Frick, M., R.~Iijima, and T.~Strzalecki} (2019): \enquote{Dynamic Random Utility,} \emph{Econometrica}, 87, 1941--2002.

\bibitem[\protect\citeauthoryear{Fudenberg and Strzalecki}{Fudenberg and Strzalecki}{2015}]{fudenberg2015dynamic}
\textsc{Fudenberg, D. and T.~Strzalecki} (2015): \enquote{Dynamic Logit with Choice Aversion,} \emph{Econometrica}, 83, 651--691.

\bibitem[\protect\citeauthoryear{Fuhrer}{Fuhrer}{2000}]{fuhrer2000habit}
\textsc{Fuhrer, J.~C.} (2000): \enquote{Habit formation in consumption and its implications for monetary-policy models,} \emph{American economic review}, 90, 367--390.

\bibitem[\protect\citeauthoryear{Gul and Pesendorfer}{Gul and Pesendorfer}{2006}]{gul2006random}
\textsc{Gul, F. and W.~Pesendorfer} (2006): \enquote{Random expected utility,} \emph{Econometrica}, 74, 121--146.

\bibitem[\protect\citeauthoryear{Hardie, Johnson, and Fader}{Hardie et~al.}{1993}]{hardie1993modeling}
\textsc{Hardie, B.~G., E.~J. Johnson, and P.~S. Fader} (1993): \enquote{Modeling loss aversion and reference dependence effects on brand choice,} \emph{Marketing science}, 12, 378--394.

\bibitem[\protect\citeauthoryear{Harmon-Jones and Mills}{Harmon-Jones and Mills}{1999}]{harmon1999cognitive}
\textsc{Harmon-Jones, E.~E. and J.~E. Mills} (1999): \emph{Cognitive dissonance: Progress on a pivotal theory in social psychology.}, American Psychological Association.

\bibitem[\protect\citeauthoryear{Honda}{Honda}{2021}]{honda2021random}
\textsc{Honda, E.} (2021): \enquote{A Model of Random Cravings,} Unpublished.

\bibitem[\protect\citeauthoryear{Hunter}{Hunter}{2005}]{hunter2005stationary}
\textsc{Hunter, J.~J.} (2005): \enquote{Stationary distributions and mean first passage times of perturbed Markov chains,} \emph{Linear Algebra and its Applications}, 410, 217--243.

\bibitem[\protect\citeauthoryear{Kashaev, Aguiar, Pl\'{a}vala, and Gauthier}{Kashaev et~al.}{2023}]{kashaev2022nonparametric}
\textsc{Kashaev, N., V.~H. Aguiar, M.~Pl\'{a}vala, and C.~Gauthier} (2023): \enquote{Dynamic and Stochastic Rational Behavior,} \emph{arXiv preprint arXiv:2302.04417}.

\bibitem[\protect\citeauthoryear{Kibris, Masatlioglu, and Suleymanov}{Kibris et~al.}{2024}]{kibrisrandom}
\textsc{Kibris, {\"O}., Y.~Masatlioglu, and E.~Suleymanov} (2024): \enquote{A random reference model,} \emph{American Economic Journal: Microeconomics}, 16, 155--209.

\bibitem[\protect\citeauthoryear{K{\H{o}}szegi and Rabin}{K{\H{o}}szegi and Rabin}{2006}]{koszegi2006model}
\textsc{K{\H{o}}szegi, B. and M.~Rabin} (2006): \enquote{A model of reference-dependent preferences,} \emph{The Quarterly Journal of Economics}, 121, 1133--1165.

\bibitem[\protect\citeauthoryear{Kovach and Suleymanov}{Kovach and Suleymanov}{2023}]{kovach2021reference}
\textsc{Kovach, M. and E.~Suleymanov} (2023): \enquote{Reference dependence and random attention,} \emph{Journal of Economic Behavior \& Organization}, 215, 421--441.

\bibitem[\protect\citeauthoryear{Li}{Li}{2023}]{li2023random}
\textsc{Li, B.} (2023): \enquote{Random utility models with status quo bias,} \emph{Journal of Mathematical Economics}, 105, 102824.

\bibitem[\protect\citeauthoryear{Li}{Li}{2022}]{li2021axiomatization}
\textsc{Li, R.} (2022): \enquote{An Axiomatization of Stochastic Utility,} \emph{arXiv preprint arXiv:2102.00143}.

\bibitem[\protect\citeauthoryear{Lu and Saito}{Lu and Saito}{2018}]{lu2018random}
\textsc{Lu, J. and K.~Saito} (2018): \enquote{Random Intertemporal Choice,} \emph{Journal of Economic Theory}, 177, 780--815.

\bibitem[\protect\citeauthoryear{Lu and Saito}{Lu and Saito}{2020}]{lu2020repeated}
---\hspace{-.1pt}---\hspace{-.1pt}--- (2020): \enquote{Repeated choice: A theory of stochastic intertemporal preferences,} Tech. rep., Working paper, Social Science Working Paper, 1449. California Institute of Technology.

\bibitem[\protect\citeauthoryear{Luce}{Luce}{1959}]{luce1959individual}
\textsc{Luce, R.~D.} (1959): \emph{Individual Choice Behavior}, John Wiley.

\bibitem[\protect\citeauthoryear{Machina}{Machina}{1985}]{machina1985stochastic}
\textsc{Machina, M.~J.} (1985): \enquote{Stochastic Choice Functions Generated From Deterministic Preferences Over Lotteries,} \emph{The Economic Journal}, 95, 575--594.

\bibitem[\protect\citeauthoryear{Masatlioglu and Ok}{Masatlioglu and Ok}{2005}]{masatlioglu2005rational}
\textsc{Masatlioglu, Y. and E.~A. Ok} (2005): \enquote{Rational choice with status quo bias,} \emph{Journal of economic theory}, 121, 1--29.

\bibitem[\protect\citeauthoryear{Milgrom and Stokey}{Milgrom and Stokey}{1982}]{milgrom1982information}
\textsc{Milgrom, P. and N.~Stokey} (1982): \enquote{Information, trade and common knowledge,} \emph{Journal of economic theory}, 26, 17--27.

\bibitem[\protect\citeauthoryear{Miller and Weinberg}{Miller and Weinberg}{2017}]{miller2017understanding}
\textsc{Miller, N.~H. and M.~C. Weinberg} (2017): \enquote{Understanding the price effects of the MillerCoors joint venture,} \emph{Econometrica}, 85, 1763--1791.

\bibitem[\protect\citeauthoryear{Morris}{Morris}{1994}]{morris1994trade}
\textsc{Morris, S.} (1994): \enquote{Trade with heterogeneous prior beliefs and asymmetric information,} \emph{Econometrica: Journal of the Econometric Society}, 1327--1347.

\bibitem[\protect\citeauthoryear{Nevo}{Nevo}{2001}]{nevo2001measuring}
\textsc{Nevo, A.} (2001): \enquote{Measuring market power in the ready-to-eat cereal industry,} \emph{Econometrica}, 69, 307--342.

\bibitem[\protect\citeauthoryear{Pennesi}{Pennesi}{2021}]{pennesi2021intertemporal}
\textsc{Pennesi, D.} (2021): \enquote{Intertemporal Discrete Choice,} \emph{Journal of Economic Behavior \& Organization}, 186, 690--706.

\bibitem[\protect\citeauthoryear{Samuelson and Zeckhauser}{Samuelson and Zeckhauser}{1988}]{samuelson1988status}
\textsc{Samuelson, W. and R.~Zeckhauser} (1988): \enquote{Status quo bias in decision making,} \emph{Journal of risk and uncertainty}, 1, 7--59.

\bibitem[\protect\citeauthoryear{Strack and Taubinsky}{Strack and Taubinsky}{2021}]{strack2021dynamic}
\textsc{Strack, P. and D.~Taubinsky} (2021): \enquote{Dynamic Preference “Reversals” and Time Inconsistency,} Tech. rep., National Bureau of Economic Research.

\bibitem[\protect\citeauthoryear{Turansick}{Turansick}{2024}]{turansick2024consumption}
\textsc{Turansick, C.} (2024): \enquote{Consumption dependent random utility,} \emph{arXiv preprint arXiv:2412.05344}.

\bibitem[\protect\citeauthoryear{Tversky and Kahneman}{Tversky and Kahneman}{1991}]{tversky1991loss}
\textsc{Tversky, A. and D.~Kahneman} (1991): \enquote{Loss aversion in riskless choice: A reference-dependent model,} \emph{The quarterly journal of economics}, 106, 1039--1061.

\bibitem[\protect\citeauthoryear{Valkanova}{Valkanova}{2020}]{valkanova2021markov}
\textsc{Valkanova, K.} (2020): \enquote{Markov stochastic choice,} Tech. rep., Mimeo.

\bibitem[\protect\citeauthoryear{Ville}{Ville}{1938}]{ville1938theorie}
\textsc{Ville, J.} (1938): \enquote{Sur la th{\'e}orie g{\'e}n{\'e}rale des jeux ou intervient l’habilet{\'e} des joueurs,} \emph{Trait{\'e} du Calcul des Probabilit{\'e}s et des ses Applications’, Paris, Gauthiers-Villars}, 171.

\end{thebibliography}

\end{document}